\documentclass[fleqn,10pt]{wlscirep}
\usepackage[utf8]{inputenc}
\usepackage[T1]{fontenc}
\usepackage{subcaption}
\usepackage{float}

\newcommand{\ti}{^\textrm{i}}
\newcommand{\tf}{^\textrm{f}}

\title{Statistical-physics-inspired model for intrinsic fluctuations driving supply and demand in markets}

\author[1]{Jan R. Mulder}
\author[1]{René van Roij}
\author[1,2, *]{Rembert A. Duine}
\affil[1]{Institute for Theoretical Physics, Utrecht University Princetonplein 5, 3584 CC, Utrecht}
\affil[2]{Department of Applied Physics, Eindhoven University of Technology, P.O. Box 513, 5600 MB Eindhoven, The Netherlands }

\affil[*]{r.a.duine@uu.nl}

\begin{abstract}
We propose a simple statistical-physics-inspired model for the effect of intrinsic fluctuations on supply and demand in markets. The model consists of agents that trade in two types of goods of which the total number is separately conserved. The relative preference of an individual agent for the two types of goods is determined by a utility that is identical for all agents. Market supply and demand curves are computed and compared for various motivated choices of the distribution of goods over the agents. In particular, we compare the ``mean-field" case, in which all agents have the same number of goods and that is akin to the economics textbook case, to the case of Boltzmann-Gibbs distributed goods, in which agents have a fluctuating number of goods. We find that the resulting equilibrium prices are not equal for these two approaches, especially when a large fraction of the agents can neither buy nor sell.

\end{abstract}
\begin{document}

\flushbottom
\maketitle
%
%
\thispagestyle{empty}

\section*{Introduction}
Consider an economy with just two agents, say Alice and Bob, who share among them two units of a certain currency in income. Disregarding any rules that may set the distribution of the income, it is twice as likely than not that the income distribution is unequal: our economy only has three ``states", one where Alice has all the income, one where Bob has everything, and one where Alice and Bob have equal income. This gives the probability $2/3$ for unequal income distributions, twice as large as the probability of $1/3$ for equal incomes. Hence, despite that Alice and Bob are each other's equals, income inequality in this simple economy emerges because of intrinsic statistical fluctuations. 

Yakovenko and co-workers proposed a similar model as the one described in our opening paragraph to compute income distributions in actual economies \cite{yakovenko}. For a large number of identical agents and fixed total amount of income, the calculation of the income distribution then becomes a textbook statistical-physics calculation \cite{blundell2010concepts}. It yields that the income distribution, i.e., the number of agents with a certain income, is an exponential function of the income with relatively many low incomes and relatively few high ones. This is the so-called Boltzmann-Gibbs distribution function. It was also found that this exponential income distribution compares favourably to economic data on actual income distributions, at least up to a certain maximum income \cite{Dr_gulescu_2001}. 

The Boltzmann-Gibbs distribution in essence results from a conservation law: any quantity that is conserved without satisfying additional constraints is exponentially distributed. One of the first occurrences of the Boltzmann-Gibbs distribution is in the theory of near-ideal gases. In that case, the probability of finding a molecule of the gas with a certain kinetic energy is proportional to the exponent of that energy divided by the temperature times Boltzmann's constant. Despite all molecules being identical, the unequal distribution of energy is understood by physicists to result from the second law of thermodynamics: the maximization of entropy. Here, the entropy is a measure of the number of possibilities to distribute the energy over the molecules and there are just many more unequal than equal states. In the economic example, the entropy is the number of ways to distribute the income over the economic agents, and is equal to 3 in the example of Alice and Bob. This definition of entropy should not be confused with other definitions of entropy in economic systems \cite{doi:10.1111/j.1749-6632.2009.05166.x}. Income inequality results because it is statistically more probable to distribute the income in an unequal way. That fluctuations may lead to inequality was also emphasized by Scheffer {\it et al.} \cite{Scheffer13154}, and puts an interesting perspective on current discussions concerning wealth and its distribution \cite{piketty,depillis}. 

Motivated by the work of Yakovenko and co-workers, we construct a simple model for the effect of intrinsic statistical fluctuations on supply and demand in markets. Our model consists of identical agents that only trade in two types of goods, where the total number of both goods is separately conserved. The supply and demand are then determined both by the utility of the individual agents and by the statistical distribution of goods. We compute and compare supply and demand curves and determine equilibrium prices for various distributions of goods. In particular, we compare the ``mean-field" situation that each agent has the same number of goods to the situation in which the goods are Boltzmann-Gibbs distributed, and find that the equilibrium prices these two approaches yield are generically different. We also discuss a possible dynamical implementation of our model.

Our motivation for constructing this model is twofold: first, our aim is to construct a model with an economic underpinning for the interactions between two agents. In physics the interactions between, say, two molecules that exchange energy, is governed by their mutual interaction potential that depends on their \emph{relative} position or momentum. The exchange of goods or money between two agents, on the other hand, is governed by the maximization of the agents' \emph{individual} utility. Interestingly, utility has, to the best of our knowledge, no counterpart in physics. Our second motivation is that, by constructing a statistical-physics based economic model with two goods of which the total number is separately conserved, one has the basis for constructing a thermodynamic description based on two variables. Based on such a thermodynamic description, it might be possible to develop a ``macroeconomic" viewpoint of our model, and it would be interesting to see if this yields anything useful, for example for the limits of the efficiency of economic cycles, akin to the Carnot efficiency in thermodynamics \cite{blundell2010concepts}. While we have not developed such a thermodynamic description yet, we present an inroad in the concluding section of this article. To conclude the discussion of our motivation we note that, while we do not expect that our model and its treatment solves open problems in present-day economics, we do believe it is interesting in its own right, in particular for econophysics researchers. 

The remainder of this article is organized as follows. We start out by introducing the set-up of our model. Hereafter, we describe the supply and demand that results for various time-independent distributions of the goods, followed by results on dynamics. We end with a brief conclusion and outlook. 

\section*{Model}
Our model consists of $N$ agents, each of which has a certain number of two goods. The distribution of goods over the agents is fully determined by $\{a_i\}$ and $\{b_i\}$, where $a_i$ and $b_i$ is the number of goods of type A and type B, respectively, that the $i$-th agent possesses. Here, $i = \left\{ 1,2,3, ...,N \right\}$ labels the agents. The total number of each type of good is separately conserved, i.e., $\sum_{i=1}^N a_i = N \bar{a}$ is conserved, with $\bar{a}$ the average number of goods of type A per agent. Similarly, $\sum_{i=1}^N b_i = N \bar{b}$, with $\bar{b}$ the average amount of goods of type B per agent, is also conserved. We take all agents to have the same utility $U (a,b)$, that determines the relative preference for the number $a$ of goods of type $A$ and the number $b$ of goods of type B. We take all trades between goods A and B to involve a fixed number $\Delta a>0$ of good A and a variable number $p>0$ of good B. The amount of goods of type B involved in the trade is the price $p$ for $\Delta a$ goods of the type A. 

The equilibrium price of $\Delta a$ goods of type A is determined by finding the intersection of the demand and supply for good A. The demand is constructed as follows. For a buying agent with $a_i$ goods of type A, and $b_i$ of type B, the utility increases in a transaction when $\Delta a\partial U (a_i, b_i)/\partial a_i  - p \partial U (a_i, b_i)/\partial b_i  >0$, where we assumed that $\Delta a \ll a_i$, and $p \ll b_i$. The demand $\tilde d_A (p)$ for good A at a price $p$ is then $\Delta a$ multiplied by the total number of agents that are able to increase their utility in trading $A$ for $B$ and that are able to pay the price $p$. (We introduce here the demand $\tilde d_A (p)$ and supply $\tilde s_A (p)$ for a given distribution of goods over the agents, that should be distinguished from the ensemble-averaged quantities $d_A (p)$ and $s_A (p)$ that are introduced below.) This yields
\begin{equation}
\label{eq:dA}
    \tilde d_A(p) = \Delta a \sum_{i=1}^N \theta\left(\frac{\partial U(a_i, b_i) / \partial a_i}{\partial U(a_i, b_i) / \partial b_i} \Delta a - p\right) \theta(b_i - p),
\end{equation}
where $\theta(x)$ is the Heaviside function: $\theta(x) = 1$ for $x >0$ and $\theta(x) = 0$, for $x < 0$. The first Heaviside function in Eq.~(\ref{eq:dA}) mathematically implements the counting of agents that are potential buyers, whereas the second one determines if they can afford the price $p$. 
Similarly, the supply $\tilde s_A (p)$ of good A is determined by 
\begin{equation}
\label{eq:sA}
    \tilde s_A(p) = \Delta a \sum_{i=1}^N \theta\left(p - \frac{\partial U(a_i, b_i) / \partial a_i}{\partial U(a_i, b_i) / \partial b_i} \Delta a \right) \theta(a_i - \Delta a),
\end{equation}
where the first factor involving the Heaviside function determines whether agent $i$ can increase its utility by selling $\Delta a$ goods of type A, whereas the second factor determines whether agent $i$ has the required $\Delta a$ goods to sell. 

While our model is formulated for an arbitrary utility $U$, we take, to be specific, the commonly-used Constant Elasticity of Substitution (CES) utility introduced by Solow in 1956 \cite{solow}:
\begin{equation}
\label{eq:CESultility}
    U(a, b) = \left[\left(\frac{a}{\alpha}\right)^r + \left(\frac{b}{\beta}\right)^r\right]^{1/r}.
\end{equation}
Here, $\alpha$ and $\beta$, or, more specific, the ratio $\beta/\alpha$, set a relative preference of the agents for goods of type A or B. The parameter $r$ sets the elasticity of substitution of the two goods \cite{solow} and will therefore influence the nature of the model.

The optimal situation for an agent, i.e. the situation in which an agent is ambiguous on whether to buy or sell good A, occurs when
\begin{equation}
    \frac{\partial U(a, b) / \partial a}{\partial U(a, b) / \partial b} \Delta a = p.
\end{equation}
Using Eq.~(\ref{eq:CESultility}) this sets the optimal ratio $ \left(b/a\right)_\textrm{opt}$ of the number of goods of type A and B desired by the agents to
\begin{equation}
    \left(\frac{b}{a}\right)_\textrm{opt} = \left(\frac{\beta}{\alpha}\right)^{r/(r-1)} \left(\frac{\Delta a}{p} \right)^{1/(r-1)} \equiv \gamma(p).
\end{equation}
This definition allows us to write Eqs.~(\ref{eq:dA}) and (\ref{eq:sA}) for the price-dependence of demand and supply of goods of type A in a more convenient way as
\begin{equation}
\begin{aligned}
\label{eq:dAsAsum}
    \tilde d_A(p) &= \Delta a \sum_{i=1}^N \theta\left[b_i - \gamma(p) a_i\right] \theta(b_i - p), \\
    \tilde s_A(p) &= \Delta a \sum_{i=1}^N \theta\left[\gamma(p) a_i - b_i \right] \theta(a_i - \Delta a).
\end{aligned}
\end{equation}
Finally, the equilibrium price $\tilde p_{\rm eq}$ of $\Delta a$ goods of the type A for a given distribution of goods $\{a_i\}$ and $\{b_i\}$ is found by solving for $\tilde p_{\rm eq}$ in the equation $\tilde s_A (\tilde p_{\rm eq}) = \tilde d_A (\tilde p_{\rm eq})$. Hence $\tilde p_{\rm eq}$ is a function of $\{a_i\}$ and $\{b_i\}$. This price should be distinguished from the average equilibrium price $p_{\rm eq}$ that we will determine below. Throughout this article, we use ``equilibrium" in the sense of economics when referring to the price, and do not use ``equilibrium" according to its physical definition in referring, e.g., to distribution functions. Hence, a non-equilibrium distribution of goods over the agents yields an equilibrium price.

\section*{Statics}
In this section, we compute ensemble-averaged supply and demand curves for two different cases and for a steady-state situation, i.e., taking the distribution of goods over the agents to be on average time-independent. Here, the ensemble-averaging corresponds to an averaging over distributions of goods over the agents. For simplicity, we assume that the probability that agent $i$ has $a_i$ goods of type A and $b_i$ goods of type $B$ is independent of the number of goods of the other agents. This implies that the probability $P(a_1, \cdots, a_N, b_1, \cdots, b_N)$ for finding the distributions of goods to be $\{ a_1, \cdots, a_N\}$ and $\{b_1, \cdots, b_N\}$ factorizes as 
\begin{equation}
    P(a_1, \cdots, a_N, b_1, \cdots, b_N) = P (a_1, b_1) P (a_2, b_2) \cdots P (a_N, b_N)~.
\end{equation}
Secondly, we assume that that quantities $\Delta a$ and price $p$ are small compared to the typical $a_i$ and $b_i$ so that we are allowed to treat $a_i$ and $b_i$ as a continuous variable. Hence, we introduce a probability density function $P(a, b)$, that determines the probability density for finding $N P (a,b)$ agents with $a$ goods of type A and $b$ of type B. Using this probability distribution, we average the demand and supply of good A in  Eq.~(\ref{eq:dAsAsum}) to obtain
\begin{equation}
\begin{aligned}
\label{eq:dAsA}
    d_A(p) &= N \Delta a \int_0^\infty da \int_0^\infty db P(a, b) \theta\left[b - \gamma(p) a \right] \theta(b - p) \equiv N \Delta a P_b, \\
    s_A(p) &= N \Delta a \int_0^\infty da \int_0^\infty db P(a, b) \theta\left[\gamma(p) a - b \right] \theta(a - \Delta a) \equiv N \Delta a P_s.
\end{aligned}
\end{equation}
In the above, we have introduced the probability $P_b$, which is the probability that an agent picked at random from all $N$ agents is a buyer at the price $p$. Similarly, the probability that an agent is a seller at the price $p$ is $P_s$. Because some agents do not have enough currency $b$ to buy, or not enough goods $a$ to sell, we have that $P_s + P_b < 1$. 

Given the average supply $s_A (p)$ and demand $d_A (p)$, the equation $d_A (p_{\rm eq}) = s_A (p_{\rm eq})$ determines the ensemble-averaged equilibrium price $p_{\rm eq}$. The fluctuations of this average price are straightforwardly computed within our earlier assumption that the agents are all uncorrelated, as in that case the number of buyers and sellers in a set of $N$ agents follows from a multinomial distribution. The variance in the demand and supply, respectively, is therefore given by
\begin{eqnarray}
\label{eq:variancesupplydemand}
\Delta_D (p) &=& N P_b (1-P_b)~,\nonumber \\
 \Delta_S (p) &=& N P_s (1-P_s)~.
\end{eqnarray}
The fluctuating demand and supply, as characterized by $\Delta_D (p)$ and $\Delta_S (p)$, will also cause the equilibrium price to fluctuate around its average value $p_{\rm eq}$. These fluctuations in the equilibrium price are now computed by solving for $p$ in the equation
\begin{equation}
    s_A (p) + \sqrt{\Delta_S (p)} = d_A (p) + \sqrt{\Delta_D (p)}~,
\end{equation}
while writing $p=p_{\rm eq}+ \Delta p$. From this we find, upon assuming $\Delta p \ll p_{\rm eq}$, that the price fluctuations $\Delta p$ are given by
\begin{equation}
\label{eq:priceflucs}
  \Delta p =\left. \frac{\left[\sqrt{\Delta_D (p)}-\sqrt{\Delta_S (p)}\right]}
  {\left[\frac{\partial s_A (p)}{\partial p} - \frac{\partial d_A (p)}{\partial p}\right]} \right|_{p=p_{\rm eq}}~.
\end{equation}
We stress that these fluctuations in the price are purely intrinsic, i.e., arise from different ways in which the goods are distributed over the agents. We have that $\Delta p \propto 1/\sqrt{N}$, so that for a large number of agents $N$ the fluctuations in the price become small. 

\begin{figure}[H]
    \centering
    \includegraphics[width=0.5\textwidth]{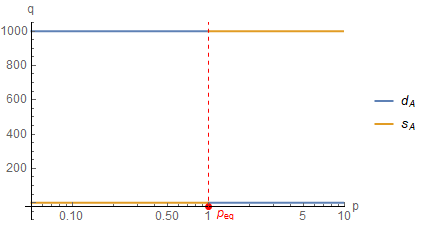}
    \caption{Demand (blue) and supply (yellow) curves in terms of quantity $q=d_A (p)$ or $q=s_A (p)$ for the mean-field situation in which each agent has the same number of goods. For simplicity we assume that here all agents have enough goods to both pay for all prices $p$, so $\bar{b}>p_{\rm eq}$, and to supply the amount $\Delta a$, i.e., $\bar{a}>\Delta a=1$ in the distribution in Eq.~(\ref{eq:pMF}), so that the curves depend only on  the number of agents and $p_{\rm MF}$ [see Eq.~(\ref{eq:pMF})] which is taken $p_{\rm MF}=1$ for this case. The number of agents is $N=1000$. }
    \label{fig:DSuniform}
\end{figure}

Let us now consider two examples. The first case we consider is that each agent has the same number of goods $\bar{a}$ of type A and $\bar{b}$ of type B. The distribution function is then
\begin{equation}
\label{eq:mfdistr}
    P_{\rm MF}(a, b) = \delta(a - \bar{a}) \delta(b - \bar{b}),
\end{equation}
which we label the ``mean-field" approximation. In the above, $\delta (\cdots)$ denotes the Dirac delta function. 
In this case, given a price $p$ either every agent wants to sell and no agent wants to buy, or vice versa. Insertion of Eq.~(\ref{eq:mfdistr}) into Eq.~(\ref{eq:dAsA}) leads to $\Delta_S (p)=\Delta_D (p)=\Delta p =0$ and hence to the step-like demand and supply curves shown in Fig.~\ref{fig:DSuniform} for $N=1000$ and equilibrium price $p_{\rm MF}$ given in Eq.~(\ref{eq:pMF}) that equals $p_{\rm MF}=1$ for the present choice of parameters. Note that we consider a logarithmic price scale. The demand and supply curves of Fig.~\ref{fig:DSuniform} show that either all agents are buyers, or they are all sellers, but there is no simultaneous supply and demand. We define the average equilibrium price to be the price at which the agents switch from demanding to supplying, which is easily established to be at $\gamma(p_{\rm MF}) = \bar{b}/\bar{a}$, which yields
\begin{equation}
\label{eq:pMF}
    \frac{p_{\textrm{MF}}}{\Delta a} \equiv \left(\frac{\bar{b}}{\bar{a}}\right)^{1-r} \left(\frac{\beta}{\alpha}\right)^r.
\end{equation}
It is, however, not clear which agents are going to sell and which agents are going to buy good A at this price. In fact, it does not matter to them if they sell or buy at price $p_{\textrm{MF}}$. Note that in this mean-field case there are no fluctuations in the price, i.e., $\Delta p=0$.  

\begin{figure}[t]
\centering
\begin{subfigure}{0.3\textwidth}
\includegraphics[width=\textwidth]{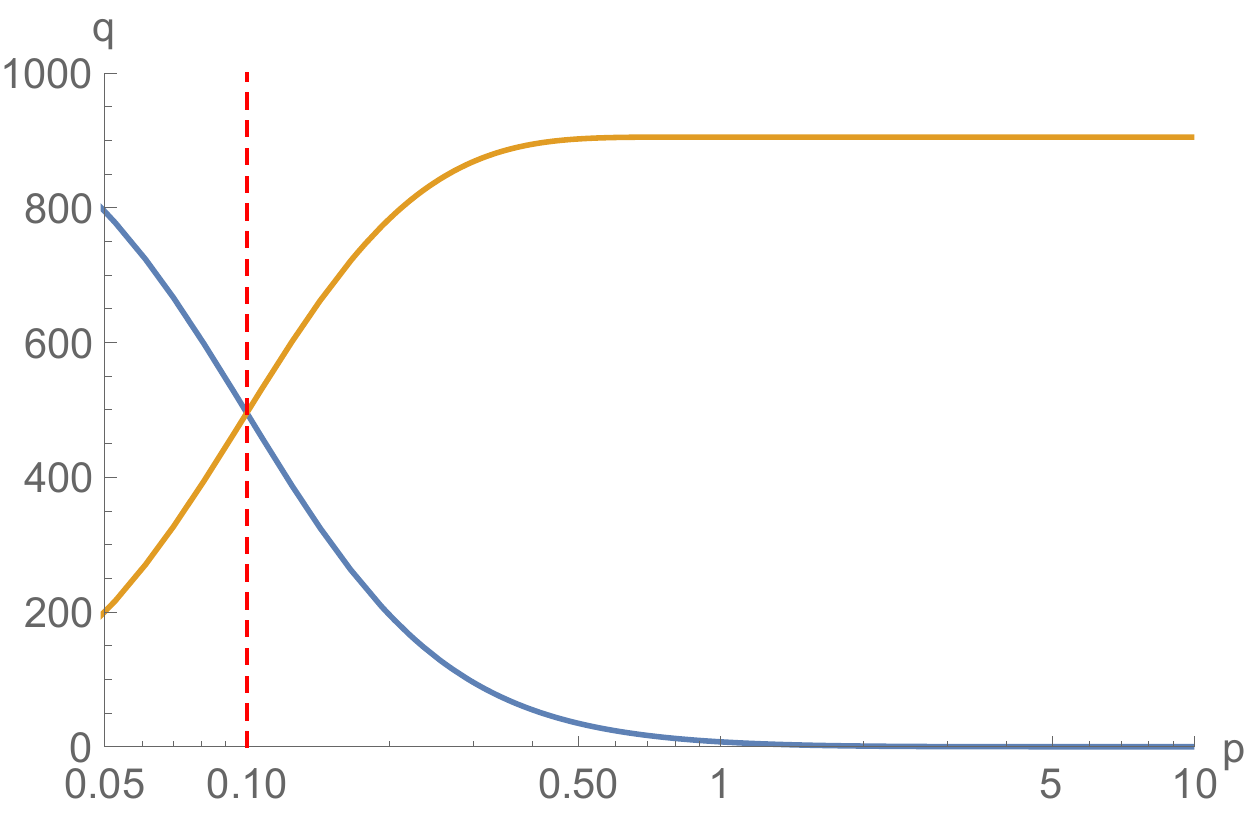}
\caption{$\beta/\alpha = 0.1$, $\bar{b} = 1$ : \\ $\Delta_{\textrm{MF}}=0$.}
\end{subfigure}
\begin{subfigure}{0.3\textwidth}
\includegraphics[width=\textwidth]{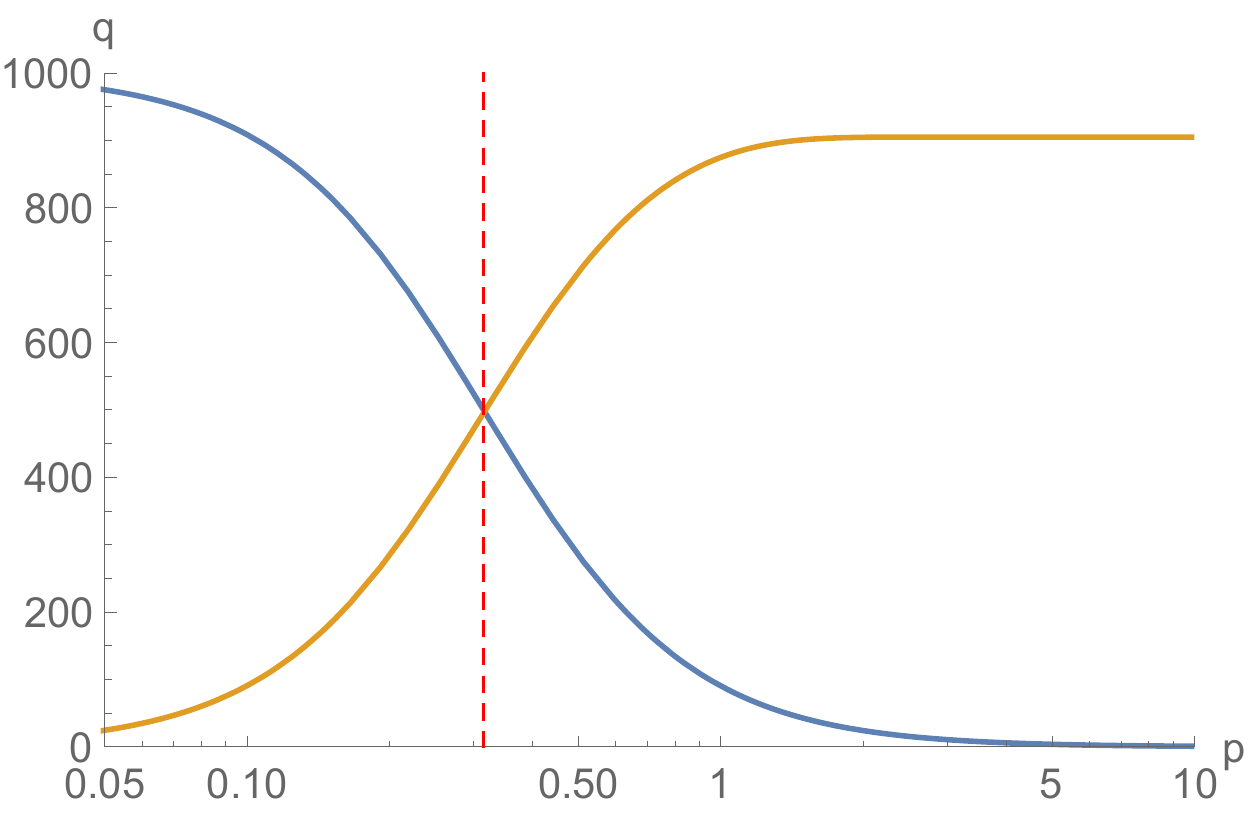}
\caption{$\beta/\alpha = 0.1$, $\bar{b} = 10$ : \\ $\Delta_{\textrm{MF}}=-0.0013$.}
\end{subfigure}
\begin{subfigure}{0.3\textwidth}
\includegraphics[width=\textwidth]{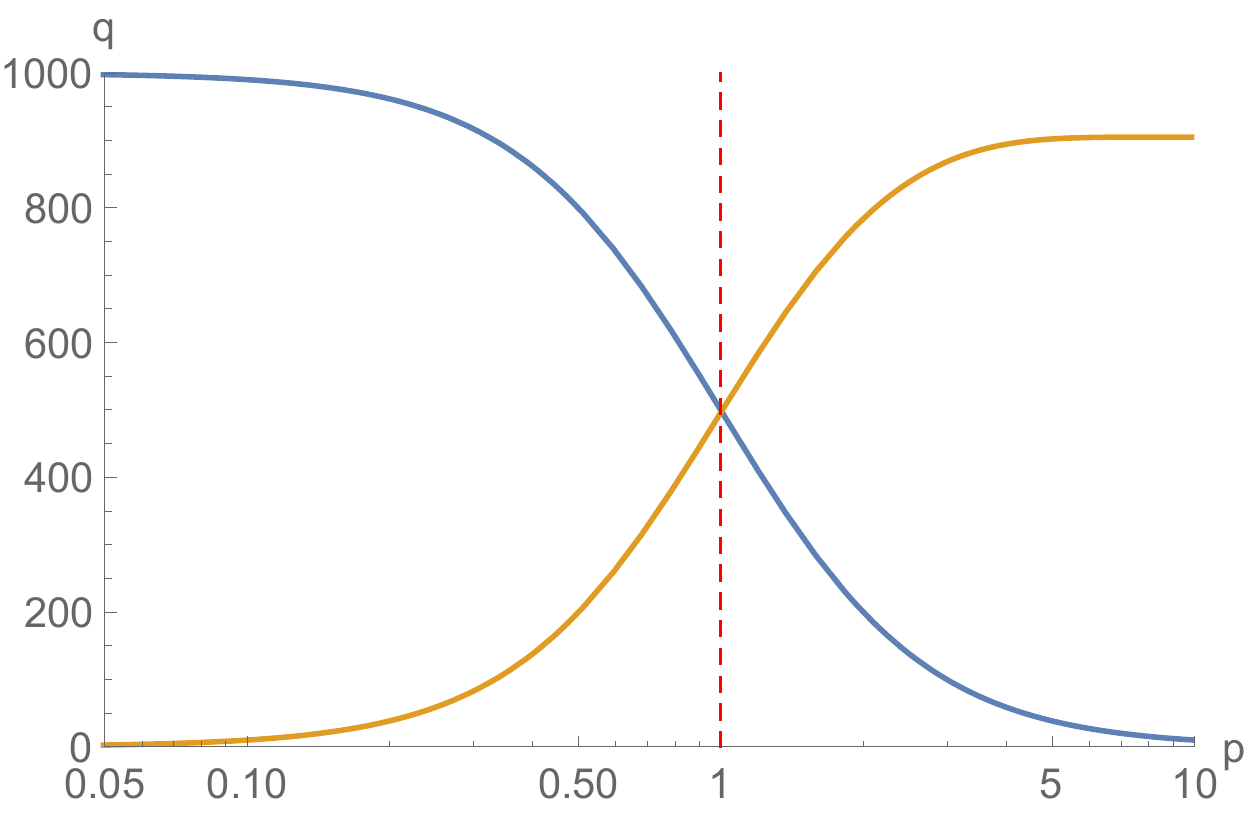}
\caption{$\beta/\alpha = 0.1$, $\bar{b} = 100$ : \\ $\Delta_{\textrm{MF}}=-0.0045$.}
\end{subfigure} \\
\begin{subfigure}{0.3\textwidth}
\includegraphics[width=\textwidth]{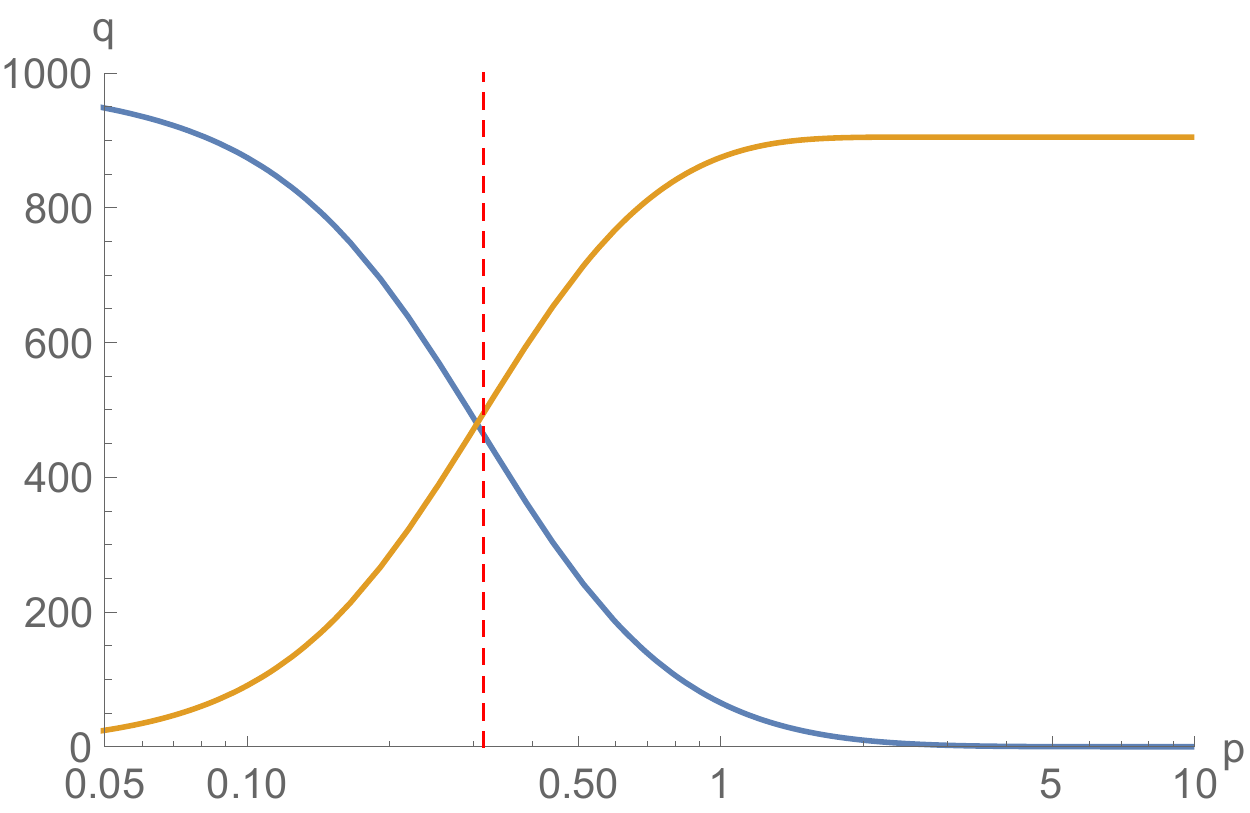}
\caption{$\beta/\alpha = 1$, $\bar{b} = 1$ : \\ $\Delta_{\textrm{MF}}=0.0102$.}
\end{subfigure}
\begin{subfigure}{0.3\textwidth}
\includegraphics[width=\textwidth]{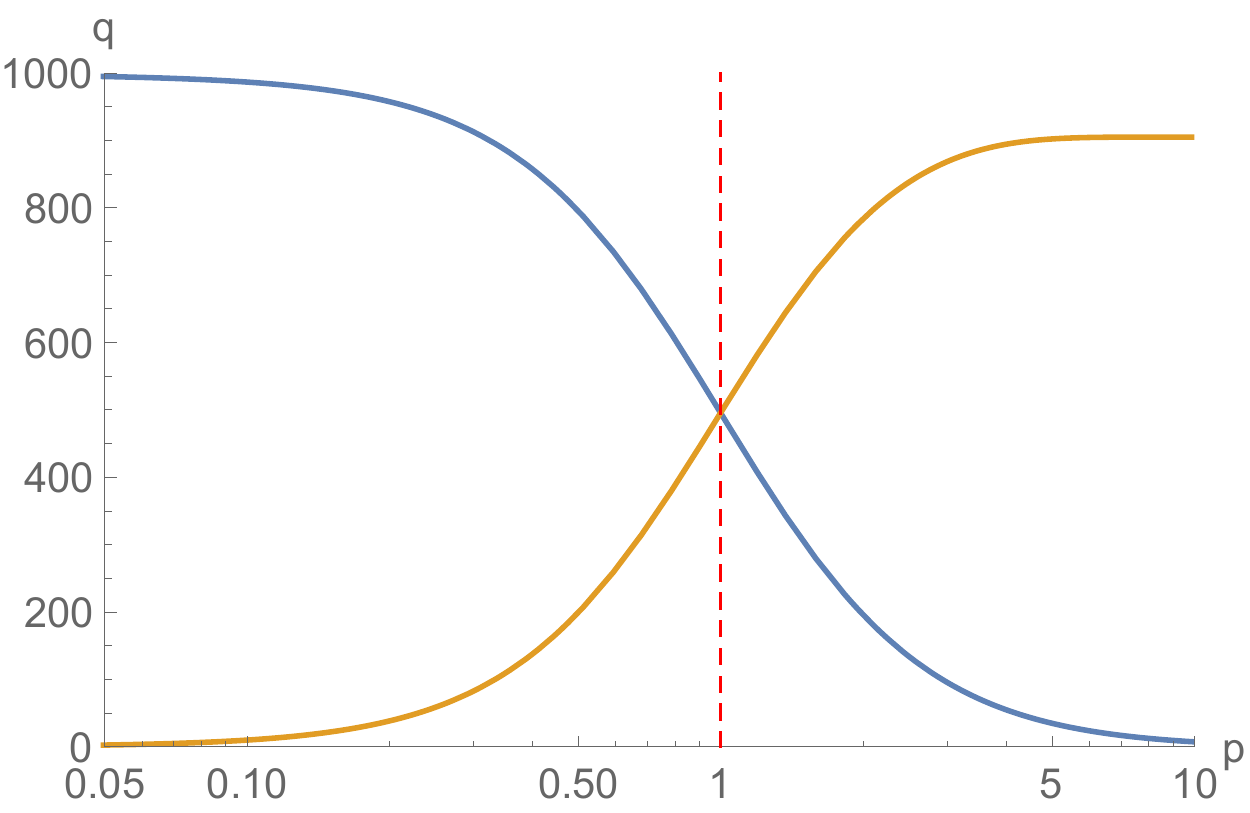}
\caption{$\beta/\alpha = 1$, $\bar{b} = 10$ : \\ $\Delta_{\textrm{MF}}=0$.}
\end{subfigure}
\begin{subfigure}{0.3\textwidth}
\includegraphics[width=\textwidth]{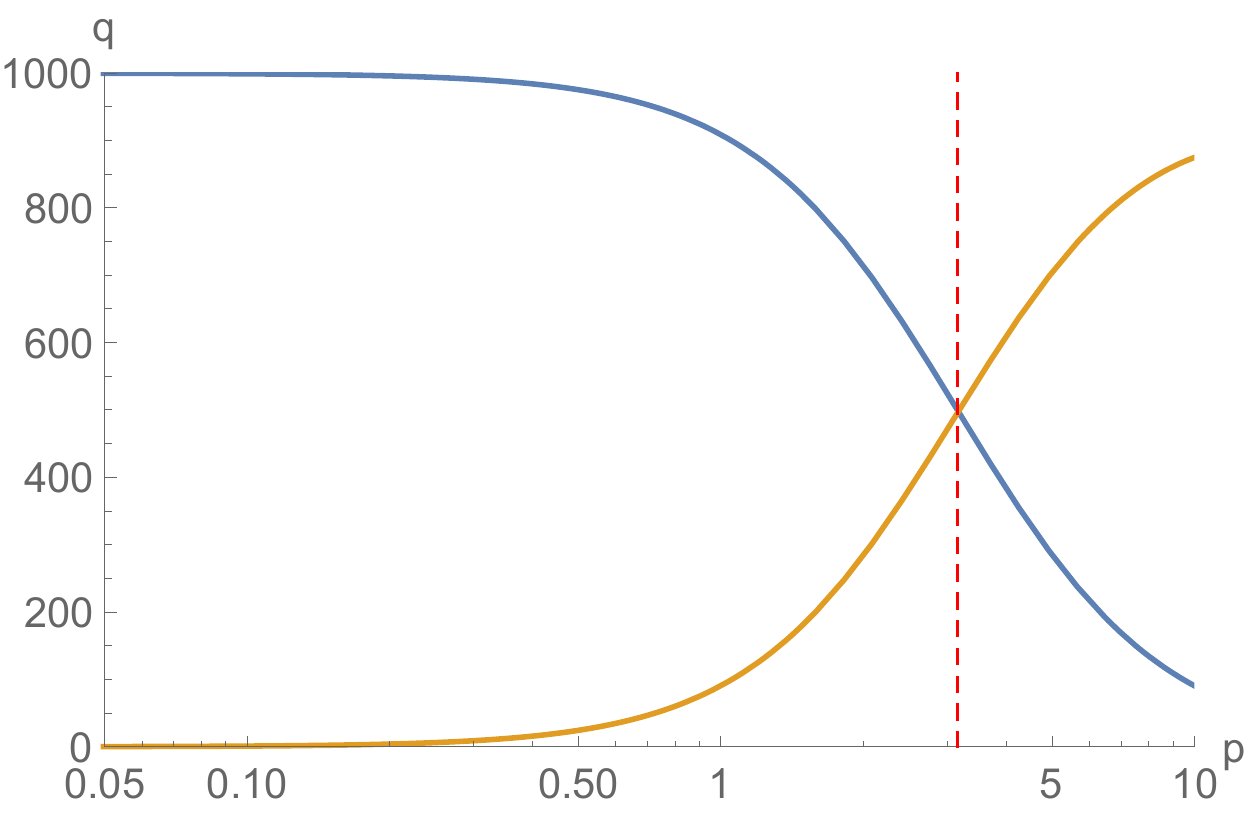}
\caption{$\beta/\alpha = 1$, $\bar{b} = 100$ : \\ $\Delta_{\textrm{MF}}=-0.0129$.}
\end{subfigure} \\
\begin{subfigure}{0.3\textwidth}
\includegraphics[width=\textwidth]{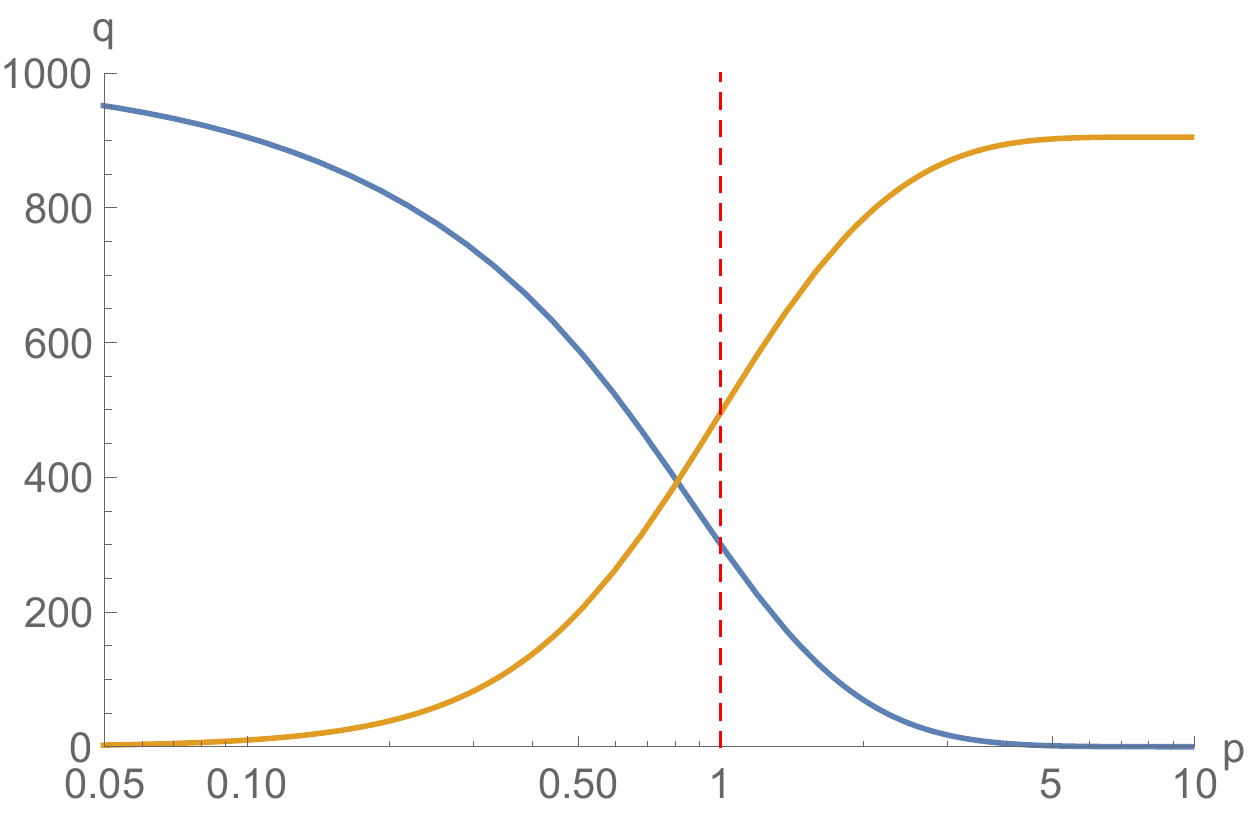}
\caption{$\beta/\alpha = 10$, $\bar{b} = 1$ : \\ $\Delta_{\textrm{MF}}=0.1896$.}
\end{subfigure}
\begin{subfigure}{0.3\textwidth}
\includegraphics[width=\textwidth]{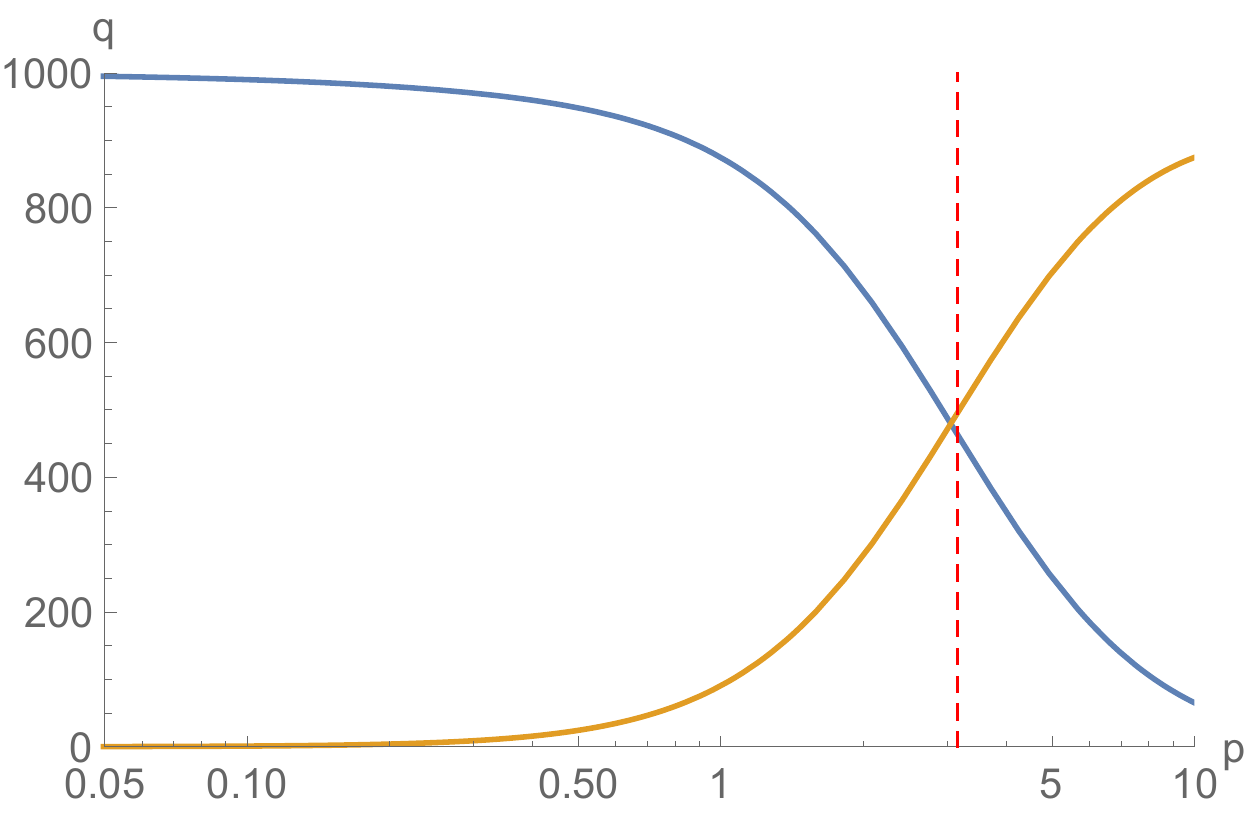}
\caption{$\beta/\alpha = 10$, $\bar{b} = 10$ : \\ $\Delta_{\textrm{MF}}=0.1012$.}
\end{subfigure}
\begin{subfigure}{0.3\textwidth}
\includegraphics[width=\textwidth]{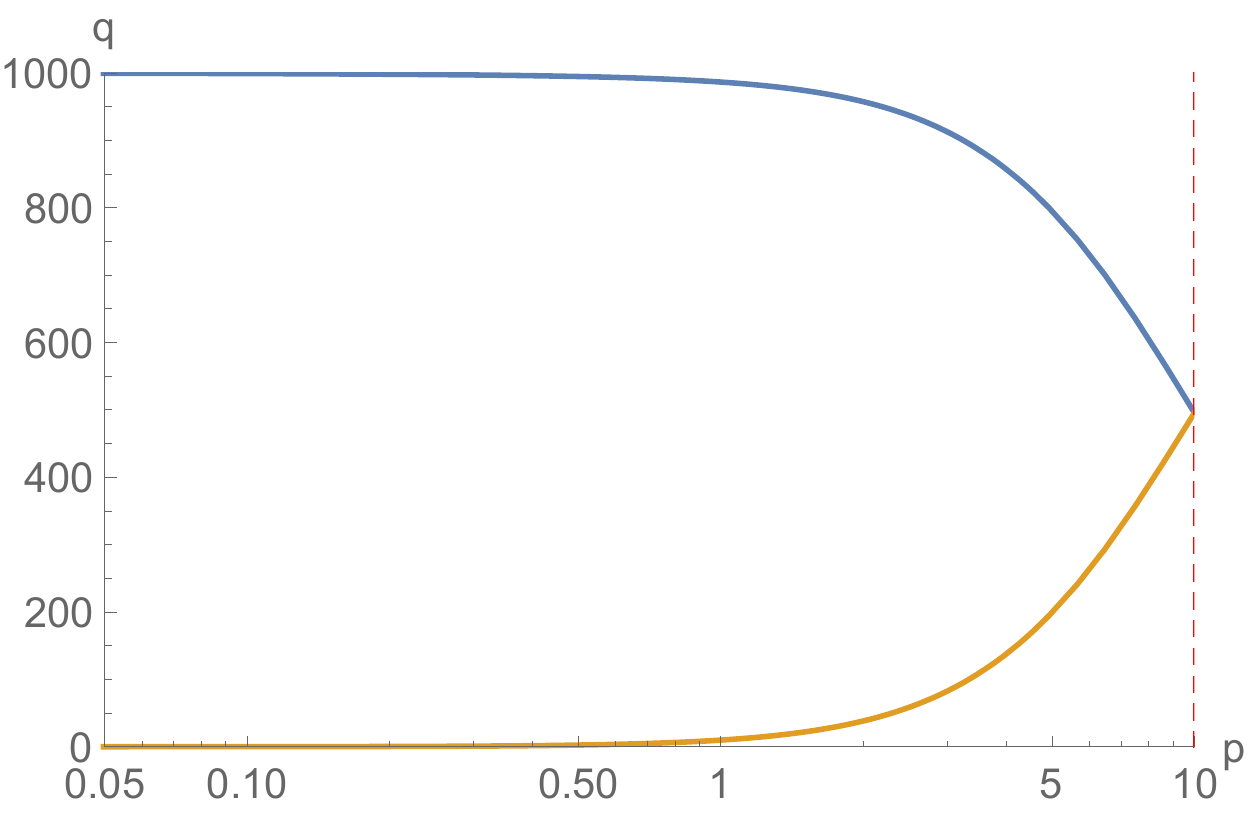}
\caption{$\beta/\alpha = 10$, $\bar{b} = 100$ : \\ $\Delta_{\textrm{MF}}=0$.}
\end{subfigure}
\caption{Demand curves (blue) and supply curves (yelow), both in terms of a quantity $q$ of goods of the type A,  plotted as a function of the price, $p$. The parameters $r = 0.5$, $\Delta a = 1$, and $\bar{a} = 10$ are kept constant. The vertical red dashed line represents the mean-field price $p_{\rm MF} = \Delta a \left(\bar{b}/\bar{a}\right)^{1-r} \left(\beta/\alpha\right)^r$ and $\Delta_{\textrm{MF}}$ is the difference between this mean-field prediction and the actual equilibrium price $p_{\rm BG}$. The number of agents $N=1000$.}
\label{fig:dAsAlog}
\end{figure}

Next, we consider a Boltzmann-Gibbs distribution of goods. This applies when the total number of goods is separately conserved and when the dynamics is such that equilibration is possible. (We return to this latter point in the next section.) This yields the distribution function,
\begin{equation}
\label{eq:BGdistr}
    P_{\rm BG}(a, b) = \frac{1}{\bar{a}\bar{b}} \exp\left(-\frac{a}{\bar{a}}\right) \exp\left(-\frac{b}{\bar{b}}\right).
\end{equation}
The exponential and factorized nature of $P_{\rm BG} (a,b)$ allows for a straightforward analytic evaluation of the integrals in Eq.~(\ref{eq:dAsA}). For the demand we find
\begin{equation}
\begin{aligned}
\label{eq:dACES}
    \frac{d^{\rm BG}_A(p)}{\Delta a} &= \frac{N}{\bar{a}\bar{b}} \int_0^\infty db\textrm{ }e^{-b/\bar{b}} \theta(b - p) \int_0^{b/\gamma(p)} da\textrm{ }e^{-a/ \bar{a}} \\
    &= \frac{N}{\bar{b}} \int_p^\infty db\textrm{ } e^{-b/\bar{b}}\left[1 - \exp\left(-\frac{b}{\gamma(p) \bar{a}}\right)\right] \\
    &= N e^{-p/\bar{b}} \left[1 - \frac{\gamma(p) \bar{a}}{\gamma(p) \bar{a} + \bar{b}} \exp\left(-\frac{p}{\gamma(p) \bar{a}}\right)\right].
\end{aligned}
\end{equation}
In a very similar way we find the supply 
\begin{equation}
\label{eq:sACES}
    \frac{s^{\rm BG}_A(p)}{\Delta a} = N e^{-\Delta a/ \bar{a}} \left[1 - \frac{\bar{b}}{\gamma(p) \bar{a} + \bar{b}} \exp\left(-\frac{\gamma(p) \Delta a}{\bar{b}}\right)\right].
\end{equation}
For a general choice of parameters, the equation $d^{\rm BG}_A(p_{\rm BG}) = s^{\rm BG}_A(p_{\rm BG})$ cannot be analytically solved for the average equilibrium price $p_{\rm BG}$ for Boltzmann-Gibbs distributed goods, i.e. with the above results for the supply and demand. In the specific case when
\begin{equation}
    \frac{\bar{a}}{\alpha} = \frac{\bar{b}}{\beta},
\end{equation}
however, the solution to $d^{\rm BG}_A(p_{\rm eq}) = s^{\rm BG}_A(p_{\rm eq})$ is actually $p_{\rm eq} = p_\textrm{MF}$. In Fig.~\ref{fig:dAsAlog}, we plot the demand and supply curves for various parameters, and compare the ``mean-field" equilibrium price to the one that results for Boltzmann-Gibbs distributed goods and that is found by numerically solving for  $p_{\rm BG}$ in $d^{\rm BG}_A(p_{\rm BG}) = s^{\rm BG}_A(p_{\rm BG})$. We find that the difference $\Delta_{\rm MF} =p_{\rm MF} - p_{\rm BG}$ between the two prices  is  generically not zero. From Fig.~\ref{fig:dAsAlog} it is also evident that in the situation that the goods are Boltzmann-Gibbs distributed, there is simultaneous supply and demand. The reason is that because of the statistical distribution of goods over all agents there are at a given price $p$ agents that are sellers and other agents that are buyers. Hence, within our model the supply and demand is  driven by statistical fluctuations. Finally, for the Boltzmann-Gibbs distribution of goods the fluctuations in the price are nonzero, $\Delta p\neq0$.

\begin{figure}[H]
\centering
\includegraphics[width=0.55\textwidth]{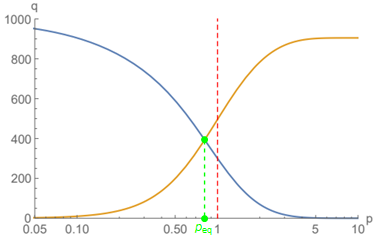}
\caption{Demand (blue) and supply (yellow) curves in terms of quantity $q$ of goods of type $A$ for $\beta/\alpha = 10$, $\bar{a} = 10$, $\bar{b} = 1$, $r = 0.5$, $\Delta a = 1$, and therefore  $p_\textrm{MF} = 1$ (indicated by the vertical red dashed line. The vertical green dashed line indicates the actual equilibrium price $p_{\rm BG} \simeq 0.8$. The number of agents $N=1000$.}
\label{fig:bigdif}
\end{figure}

To get a better understanding of the difference between the equilibrium price for the mean-field case and the Boltzmann-Gibbs case, we take a closer look at a the specific set of parameters of Fig.~\ref{fig:dAsAlog}~(g) in which there is a relatively large difference between the two prices. This situation is presented in Fig.~\ref{fig:bigdif}, and corresponds to the parameters  $\beta/\alpha = 10$, $\bar{a} = 10$, $\bar{b} = 1$, $r = 0.5$, $\Delta a = 1$. For these parameters, the  mean-field price for good A is $1$ unit of good B. However, on average agents have only $\bar{b} = 1$ of good B. This means that, for Boltzmann-Gibbs distributed goods, that $\int_0^1 db \exp(-b) \approx 0.63$ of the agents cannot afford to buy good A at a price that is near the mean-field price, and are therefore not a buyer. Furthermore, around $\int_0^1 da \exp(-a/10)/10 \approx 0.1$ of the agents do not have enough of good A to sell $\Delta a$, and are therefore not a seller. So, statistically speaking, about $6\%$ of the agents is neither a buyer nor a seller. This means that these agents do not contribute to the demand or the supply curves around the mean-field price. For the other parameters in Fig.~\ref{fig:dAsAlog} the group of agents that are neither a buyer nor a seller is significantly smaller, making the mean-field price a better approximation to the actual price. Furthermore, in the cases where $\bar{a}/\alpha = \bar{b}/\beta$, an equivalent amount of good A and good B is taken out of play, leading to the exact same equilibrium price, as predicted. In view of this, we conclude that the deviations of the equilibrium price for a Boltzmann-Gibbs distribution of goods from the mean-field price are largest when the fraction of agents that cannot sell or buy are relatively large, and when these fraction are unequal. We also find that the mean-field price can be larger or smaller than the price for the Boltzmann-Gibbs distribution, although not in a symmetric fashion as a function of the various parameters.

\section*{Dynamics}
The Boltzmann-Gibbs distribution that we assumed in the previous section is understood as the steady-state distribution that arises when goods of the type A and B are exchanged randomly in a way that obeys detailed balance, similar to the redistribution of income in the model of Yakovenko and co-workers \cite{yakovenko}. This dynamics could be understood as effective dynamics that arises because of the presence of many other goods and commodities that may be traded with goods A and B, and of which the dynamics is --- to use physics language --- ``integrated out". In this section we look into the opposite limit in which the dynamics is fully determined by the utility $U (a,b)$.

A single time step of the dynamics that we consider is as follows. First, the supply and demand is computed from the distribution of goods cf. Eqs.~(\ref{eq:dA})~and~(\ref{eq:sA}), which leads to the equilibrium price $\tilde p_{\rm eq}$ that is found from numerically solving $\tilde s_A (\tilde p_{\rm eq}) = \tilde s_ B (\tilde p_{\rm eq})$. At the equilibrium price, the number of selling agents by construction equals the number of buying agents. Therefore, all agents that want to buy or sell can do so. Subsequently, the agents are randomly paired up in pairs of one buyer and one seller, and all agents trade $\Delta a$ goods of type A for $\tilde p_{\rm eq}$ goods of type B.

We consider two initial distributions of goods. The first one is the Boltzmann-Gibbs distribution $P_{\rm BG} (a,b)$ of Eq.~(\ref{eq:BGdistr}). The second one is a uniform distribution of goods of type A on the interval $[0, 2\bar{a}]$, and a uniform distribution of goods of type B on the interval $[0, 2\bar{b}]$. This distribution is mathematically given by  
\begin{equation}
\label{eq:PUnif}
    P_{\rm un} (a,b) = \frac{1}{4 \bar{a} \bar{b}}\theta(a)\theta(2\bar{a}-a)\theta(b)\theta(2\bar{b}-b).
\end{equation}

\begin{figure}[p]
    \centering
    \begin{subfigure}{0.3\textwidth}
    \includegraphics[width=\textwidth]{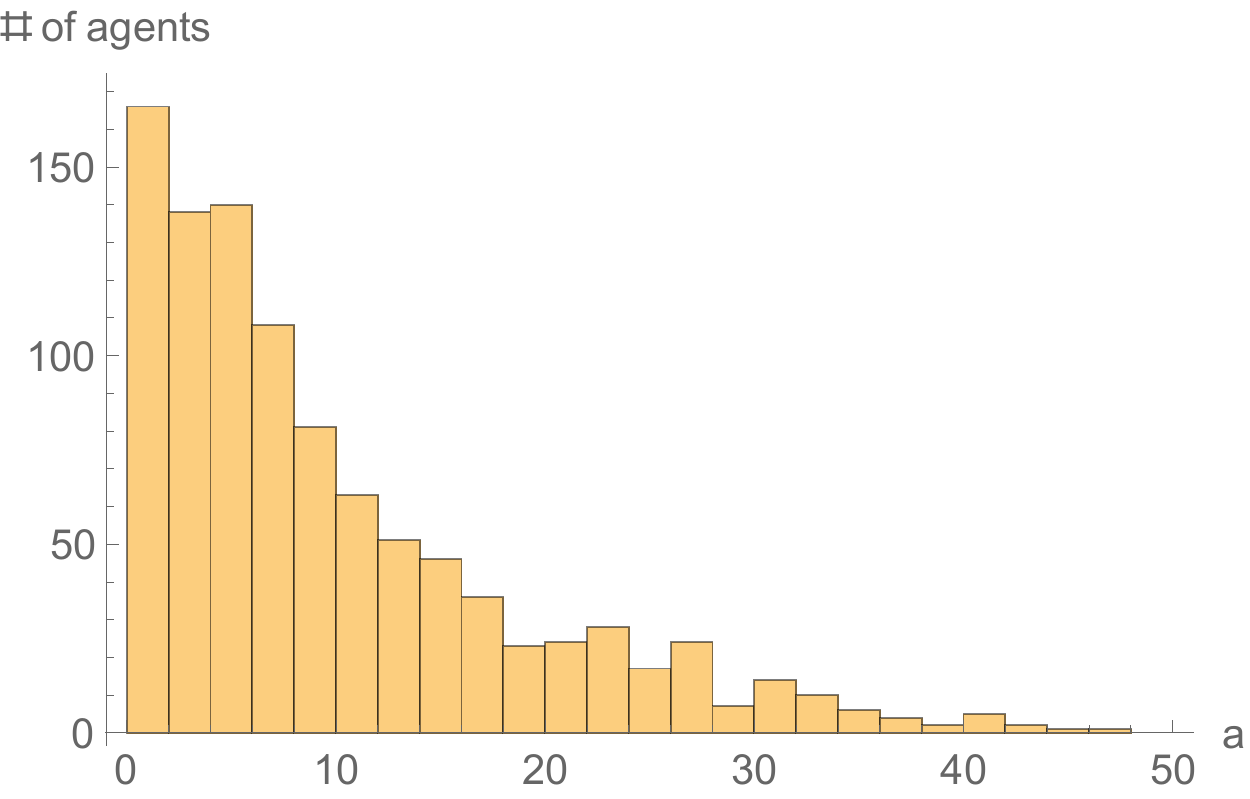}
    \caption{Distribution of A at $t=0$.}
    \end{subfigure}
    \begin{subfigure}{0.3\textwidth}
    \includegraphics[width=\textwidth]{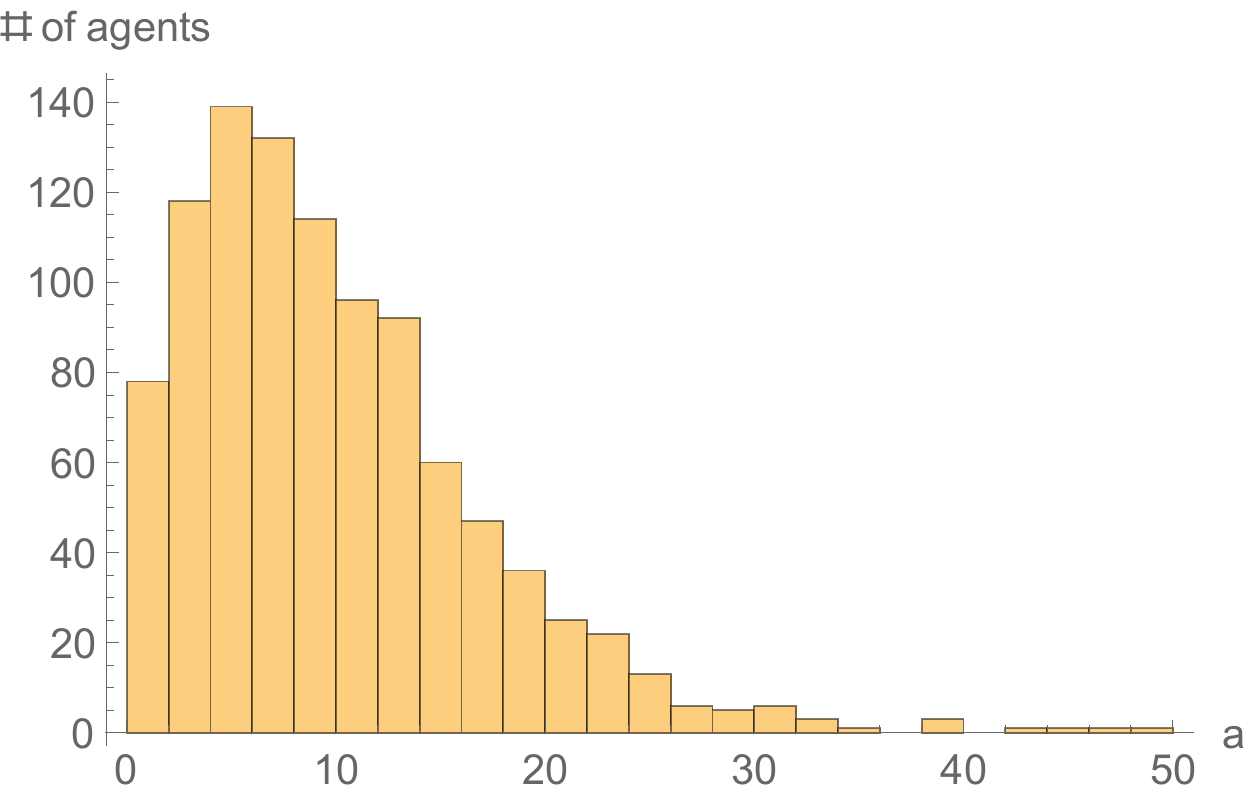}
    \caption{Distribution of A at $t=2000$.}
    \end{subfigure} \\
    \begin{subfigure}{0.3\textwidth}
    \includegraphics[width=\textwidth]{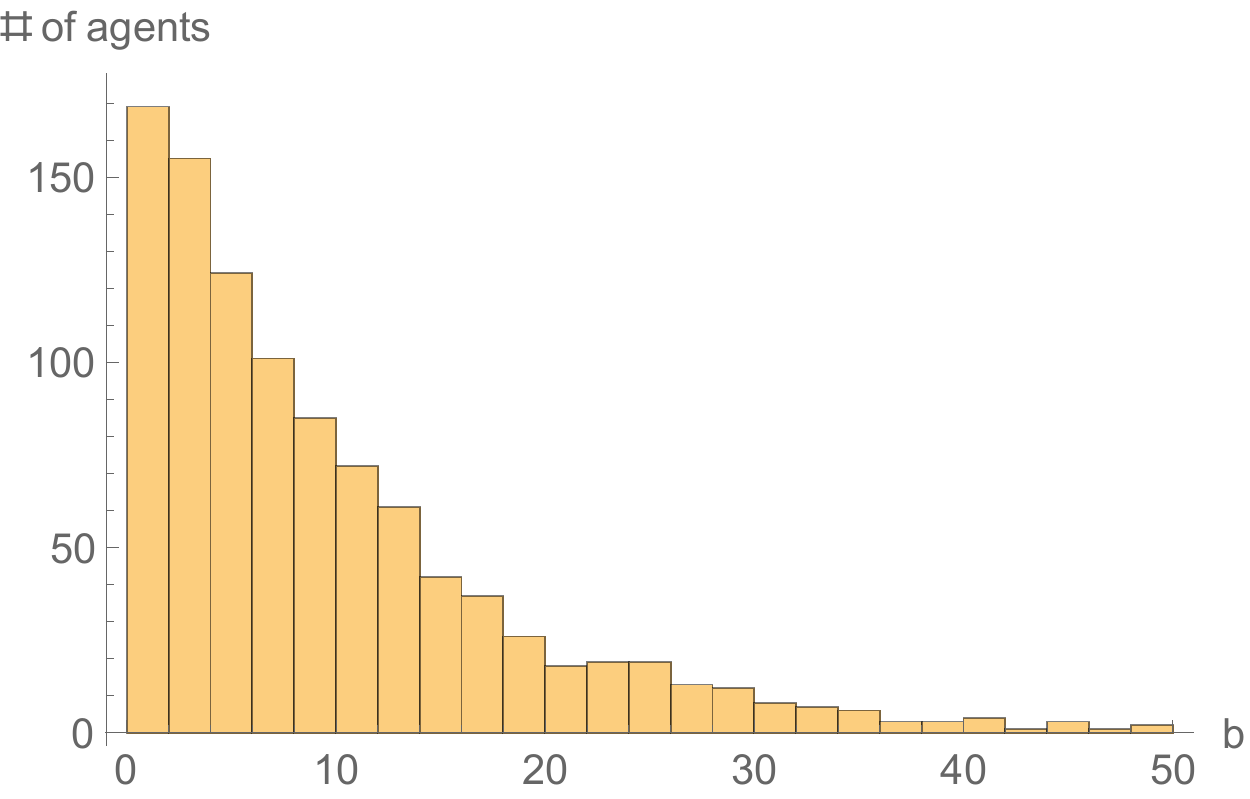}
    \caption{Distribution of B at $t=0$.}
    \end{subfigure}
    \begin{subfigure}{0.3\textwidth}
    \includegraphics[width=\textwidth]{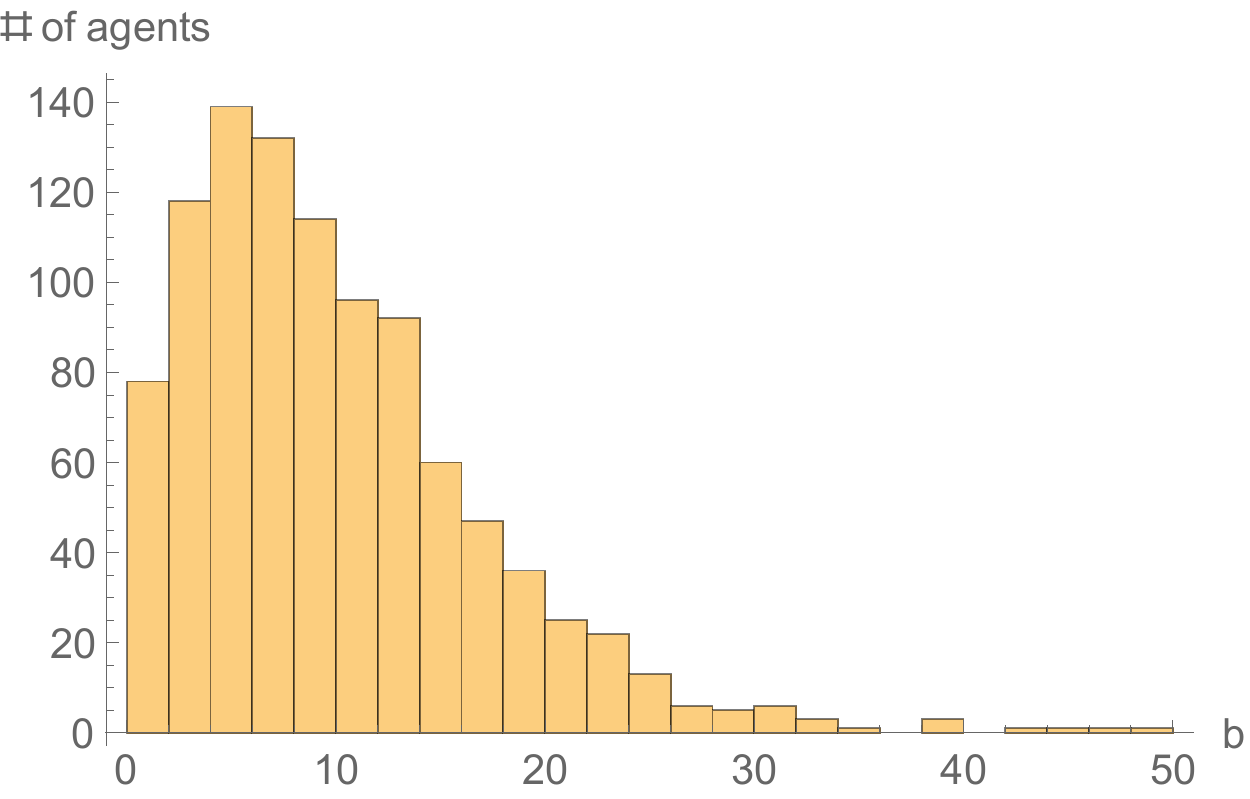}
    \caption{Distribution of B at $t=2000$.}
    \end{subfigure} \\
    \begin{subfigure}{0.3\textwidth}
    \includegraphics[width=\textwidth]{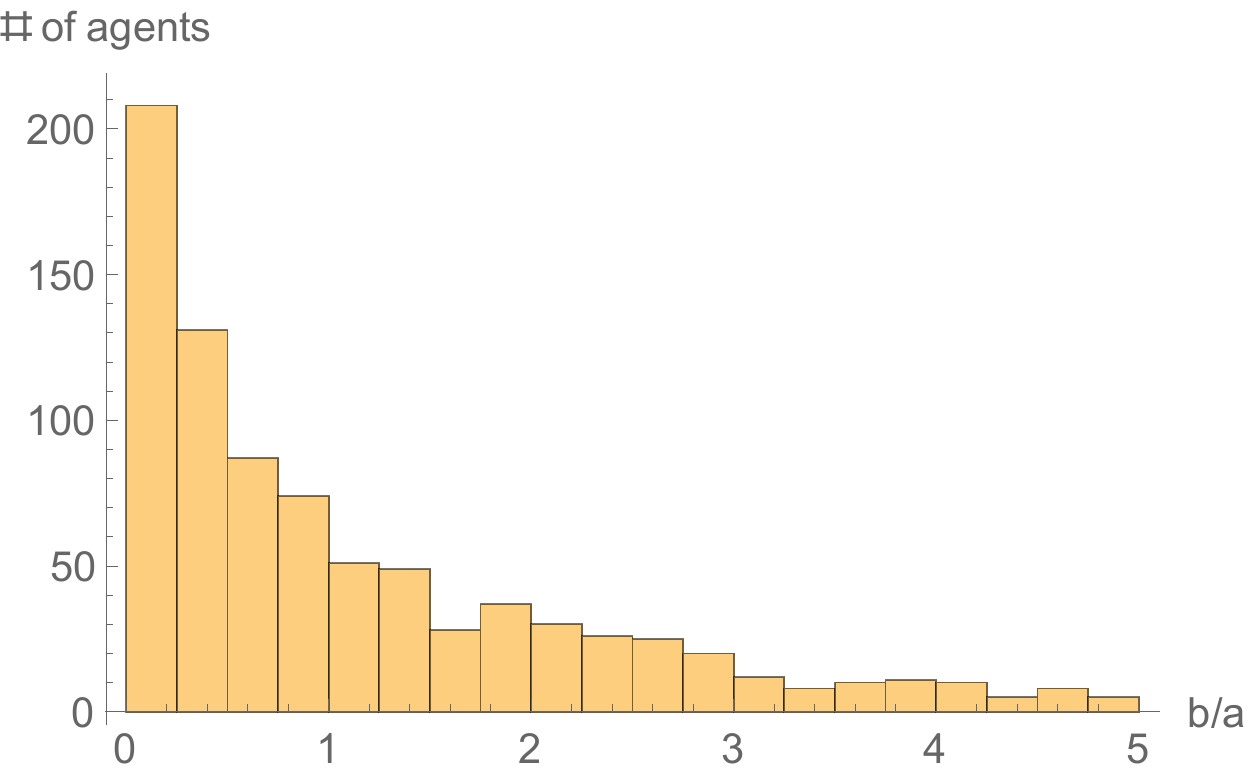}
    \caption{Distribution of $b/a$ at $t=0$.}
    \end{subfigure}
    \begin{subfigure}{0.3\textwidth}
    \includegraphics[width=\textwidth]{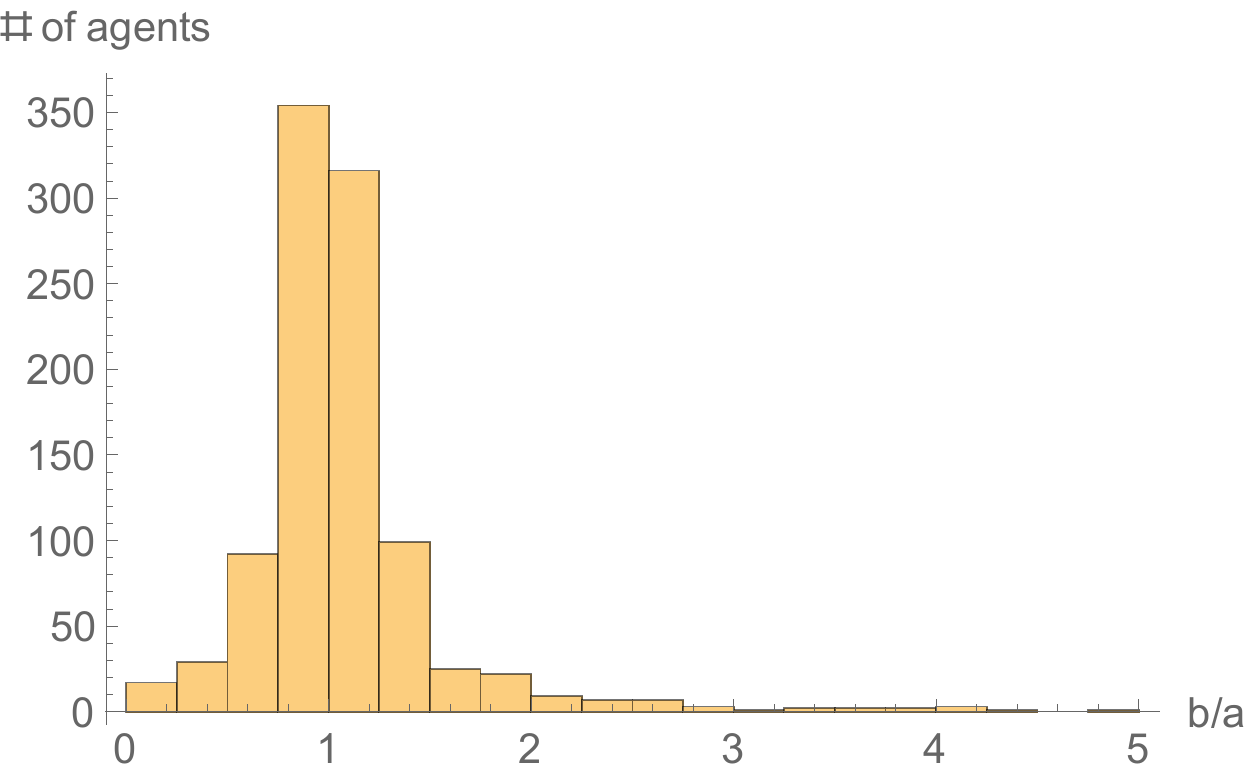}
    \caption{Distribution of $b/a$ at $t=2000$.}
    \end{subfigure} \\
    \begin{subfigure}{0.3\textwidth}
    \includegraphics[width=\textwidth]{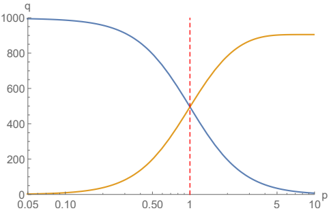}
    \caption{(blue) Demand and (yellow) supply curves at $t=0$.}
    \end{subfigure}
    \begin{subfigure}{0.3\textwidth}
    \includegraphics[width=\textwidth]{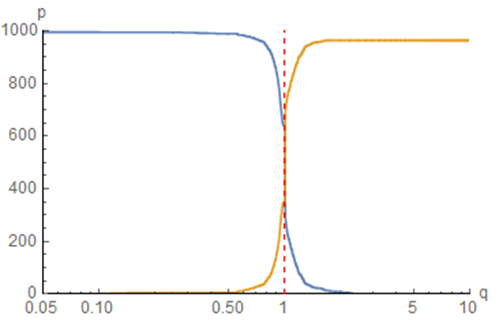}
    \caption{(blue) Demand and (yellow) supply curves at $t=2000$.}
    \end{subfigure}
    \caption{Evolution of system due to market trades with initial Boltzmann distribution, for the parameters $\bar{a} = \bar{b} = 10$, $\alpha = \beta = 1$, $r = 0.5$, and $\Delta a = 1$. The number of agents $N=1000$. Shown are the distributions for goods and their ratio, and the initial supply and demand curves, as well as the supply and demand after $t=2000$ simulation steps which we checked to be sufficient to reach a steady state.}
    \label{fig:EvoBoltz}
\end{figure}

\begin{figure}[p]
    \centering
    \begin{subfigure}{0.3\textwidth}
    \includegraphics[width=\textwidth]{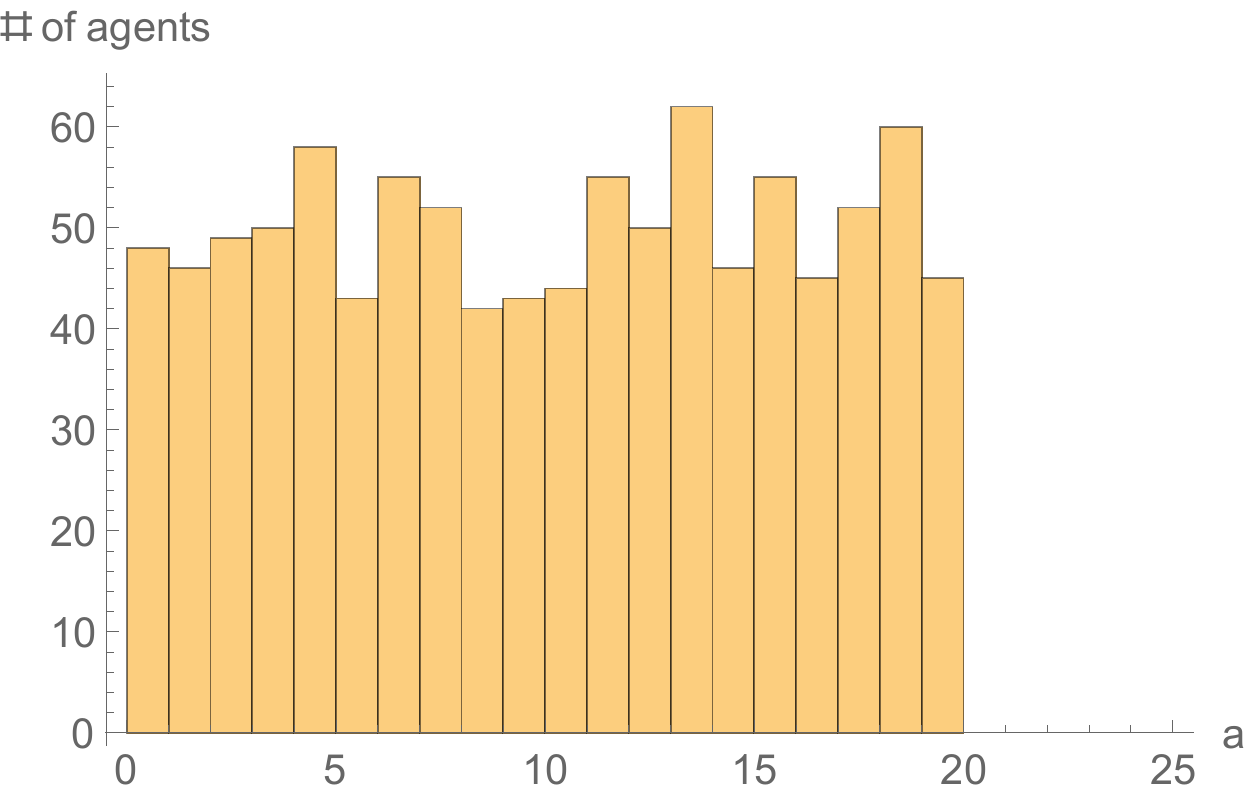}
    \caption{Distribution of A at $t=0$.}
    \end{subfigure}
    \begin{subfigure}{0.3\textwidth}
    \includegraphics[width=\textwidth]{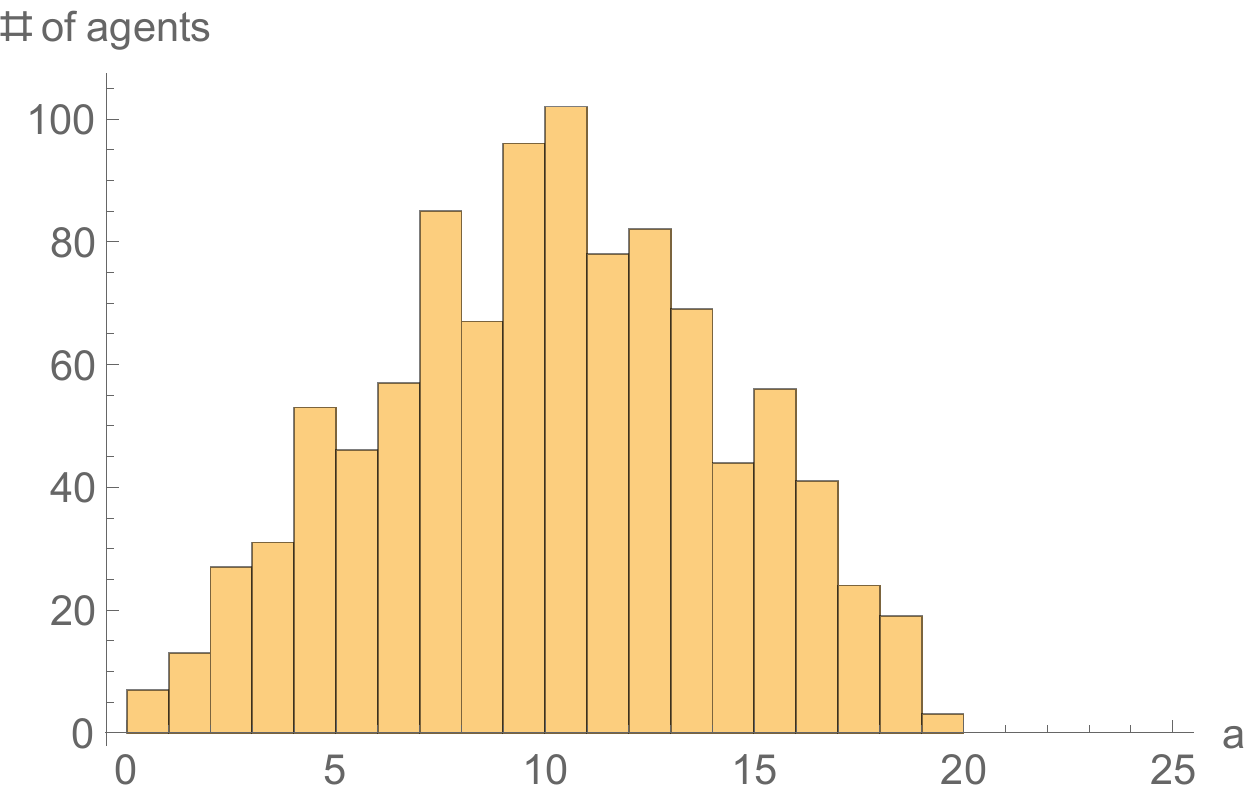}
    \caption{Distribution of A at $t=2000$.}
    \end{subfigure} \\
    \begin{subfigure}{0.3\textwidth}
    \includegraphics[width=\textwidth]{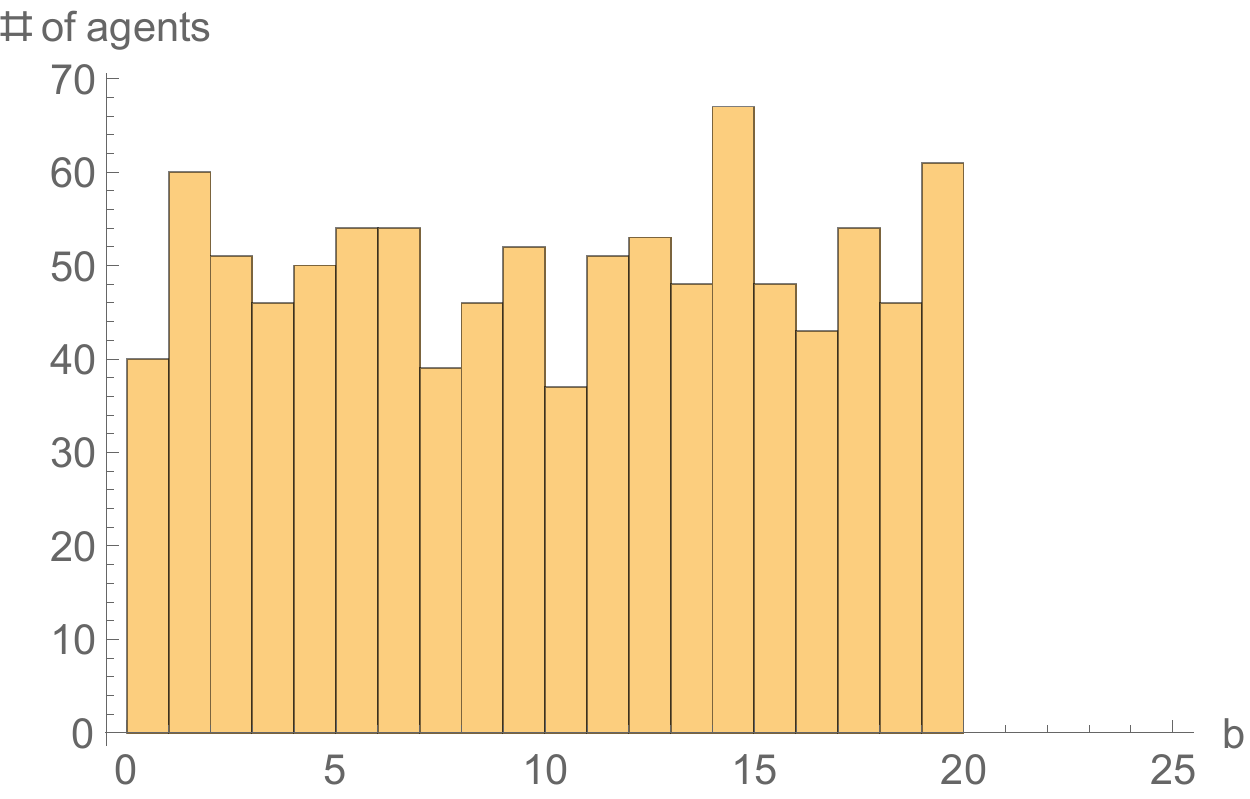}
    \caption{Distribution of B at $t=0$.}
    \end{subfigure}
    \begin{subfigure}{0.3\textwidth}
    \includegraphics[width=\textwidth]{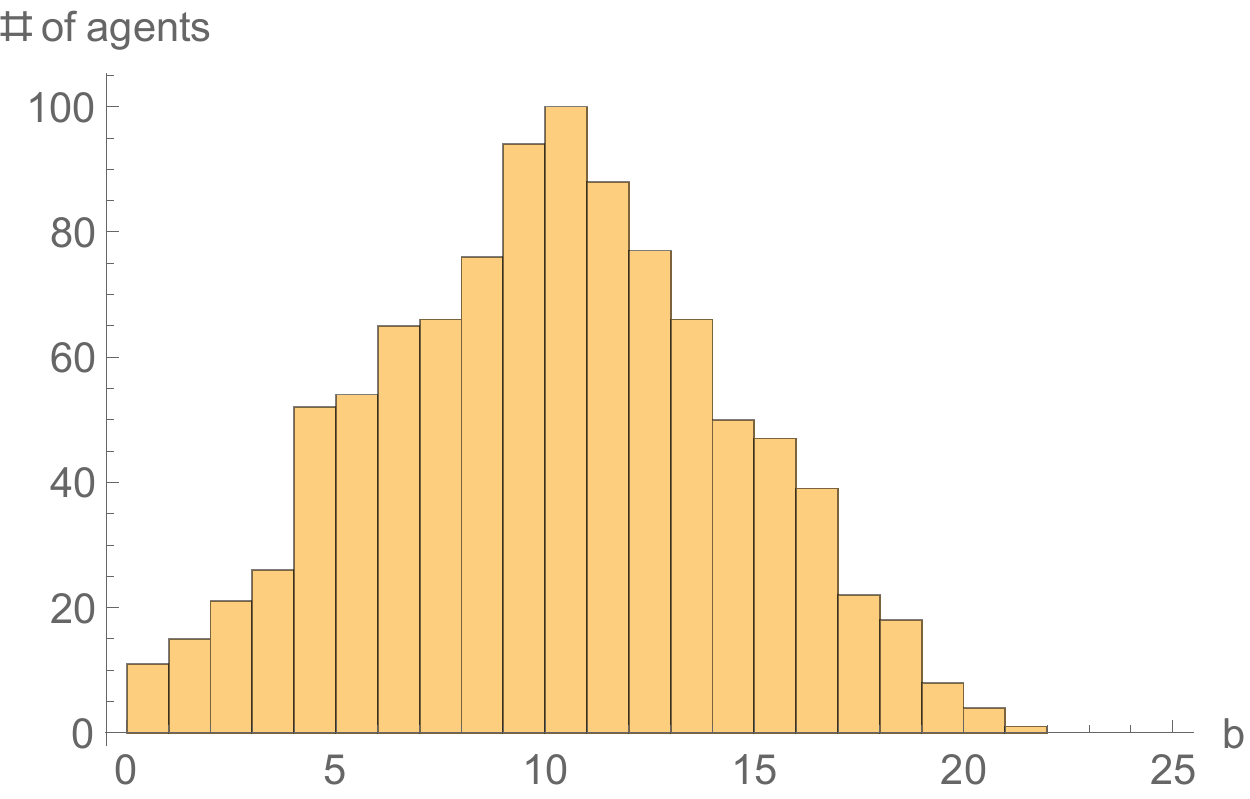}
    \caption{Distribution of B at $t=2000$.}
    \end{subfigure} \\
    \begin{subfigure}{0.3\textwidth}
    \includegraphics[width=\textwidth]{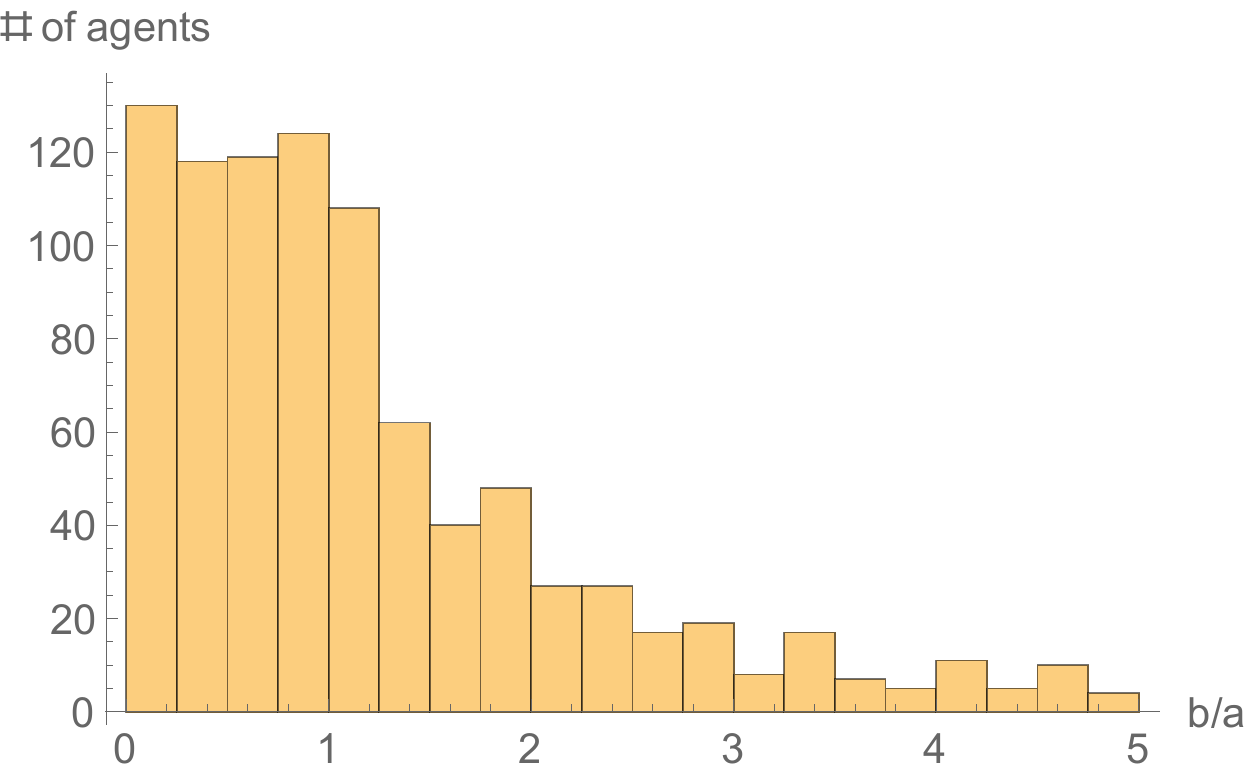}
    \caption{Distribution of $b/a$ at $t=0$.}
    \end{subfigure}
    \begin{subfigure}{0.3\textwidth}
    \includegraphics[width=\textwidth]{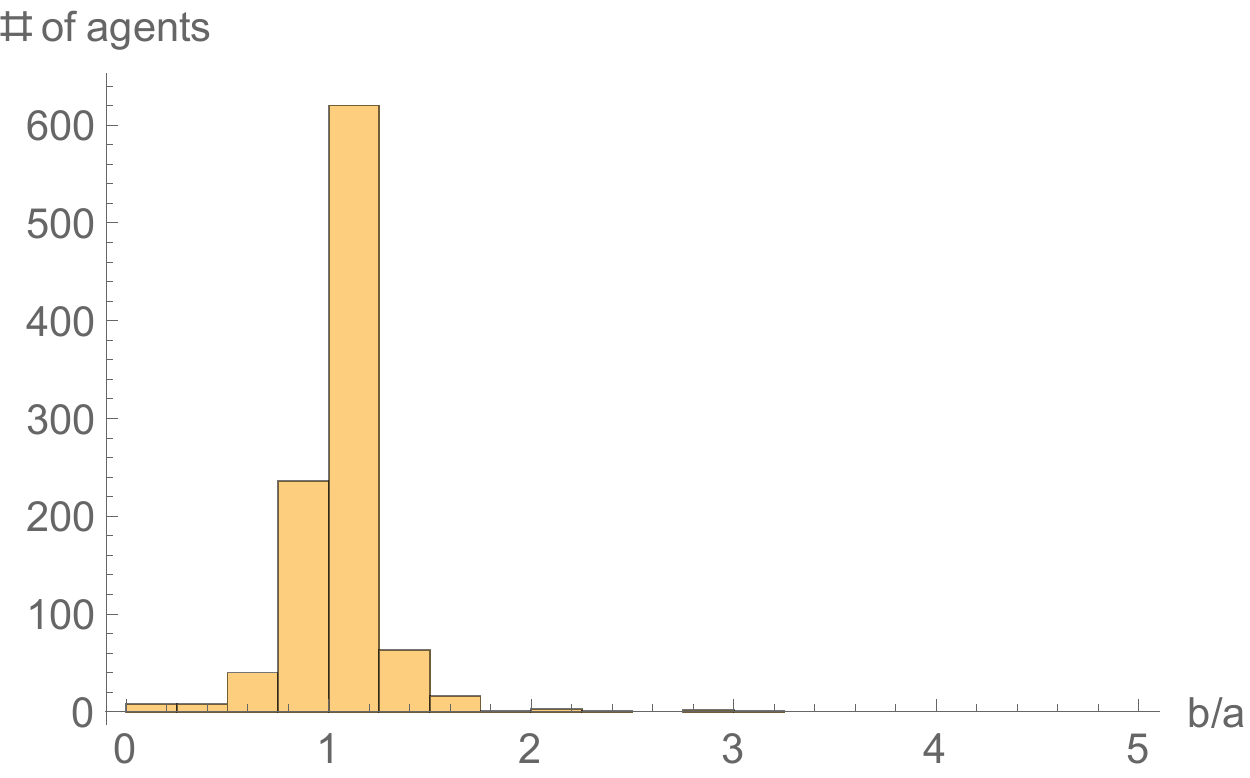}
    \caption{Distribution of $b/a$ at $t=2000$.}
    \end{subfigure} \\
    \begin{subfigure}{0.3\textwidth}
    \includegraphics[width=\textwidth]{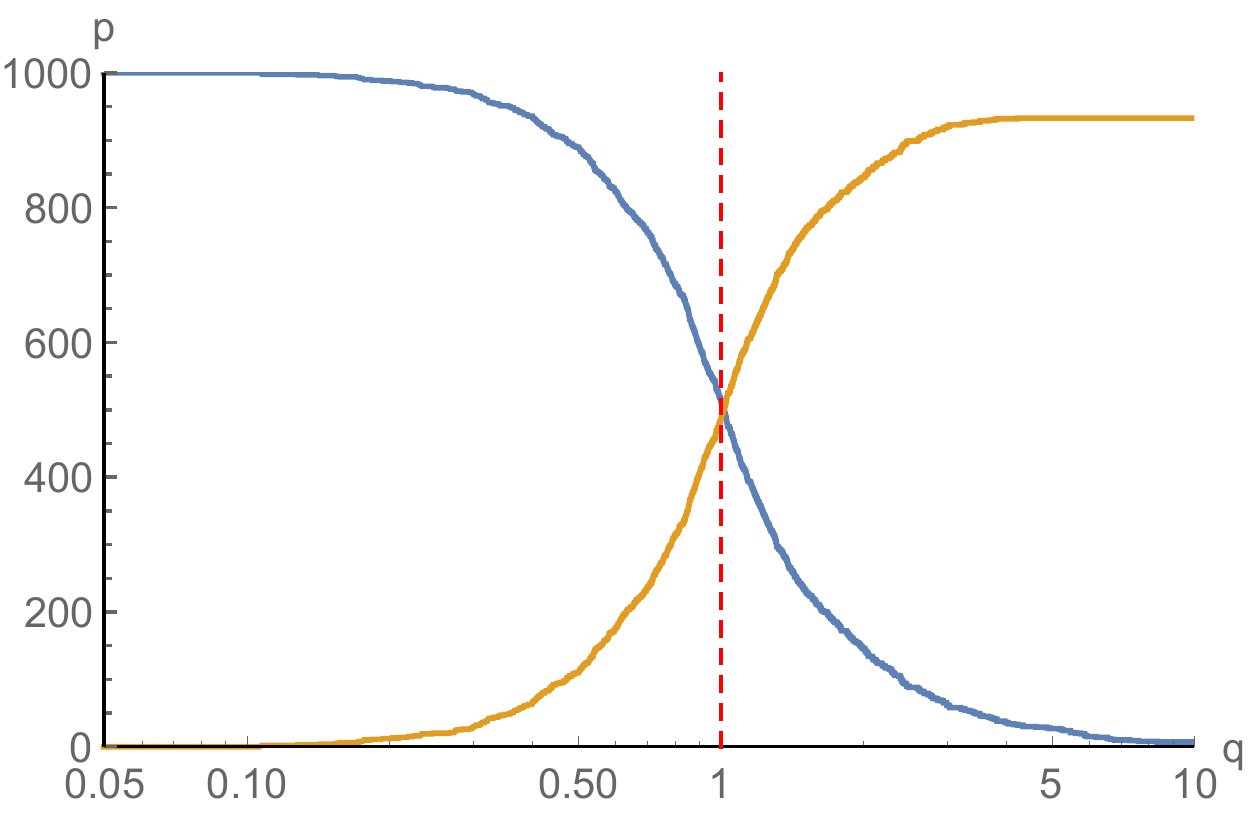}
    \caption{(blue) Demand and (yellow) supply curves at $t=0$.}
    \end{subfigure}
    \begin{subfigure}{0.3\textwidth}
    \includegraphics[width=\textwidth]{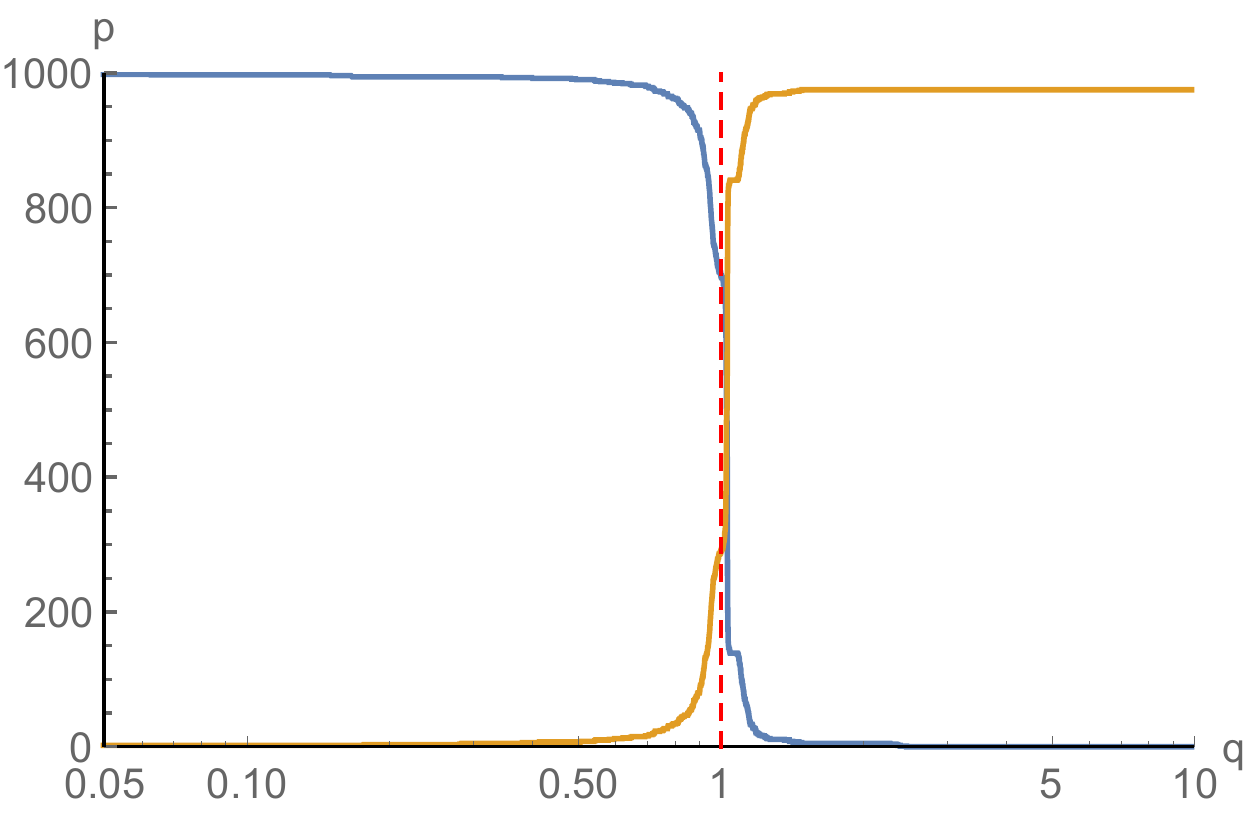}
    \caption{(blue) Demand and (yellow) supply curves at $t=2000$.}
    \end{subfigure}
    \caption{Evolution of system due to market trades with initial uniform distribution, for the parameters $\bar{a} = \bar{b} = 10$, $\alpha = \beta = 1$, $r = 0.5$, and $\Delta a = 1$. The number of agents $N=1000$. Shown are the distributions for goods and their ratio, and the initial supply and demand curves, as well as the supply and demand after $t=2000$ simulation steps.}
    \label{fig:EvoUnif}
\end{figure}

We have performed simulations to find the time evolution of the distribution of goods over the agents for these two initial distributions of goods. In the simulations, representations of the initial Boltzmann-Gibbs distribution are obtained by letting the system ``thermalize" by randomly exchanging a fixed small number of goods between agents. A representation of the initial uniform distribution is set by giving each agent a random amount of goods chosen from the intervals $[0, 2\bar{a}]$ and  $[0, 2\bar{b}]$ for goods A and B, respectively. The results of our simulations for the initial Boltzmann-Gibbs and uniform distributions are shown in Fig.~\ref{fig:EvoBoltz} and \ref{fig:EvoUnif}, respectively, for $N=1000$ and for the parameters $\bar{a} = \bar{b} = 10$, $\alpha = \beta = 1$, $r = 0.5$, and $\Delta a = 1$. For these parameters the average number of goods of type A is equal to average number of goods of the type B. Furthermore, agents get equal utility from good A and good B (no preference). Therefore, the mean-field price $p_{\textrm{MF}} = 1$ [see Eq.~(\ref{eq:pMF})], and the preferred ratio of goods is one-to-one, i.e., $(b/a)_{\rm opt}=1$.


\begin{figure}[H]
    \centering
    \includegraphics[width=0.5\textwidth]{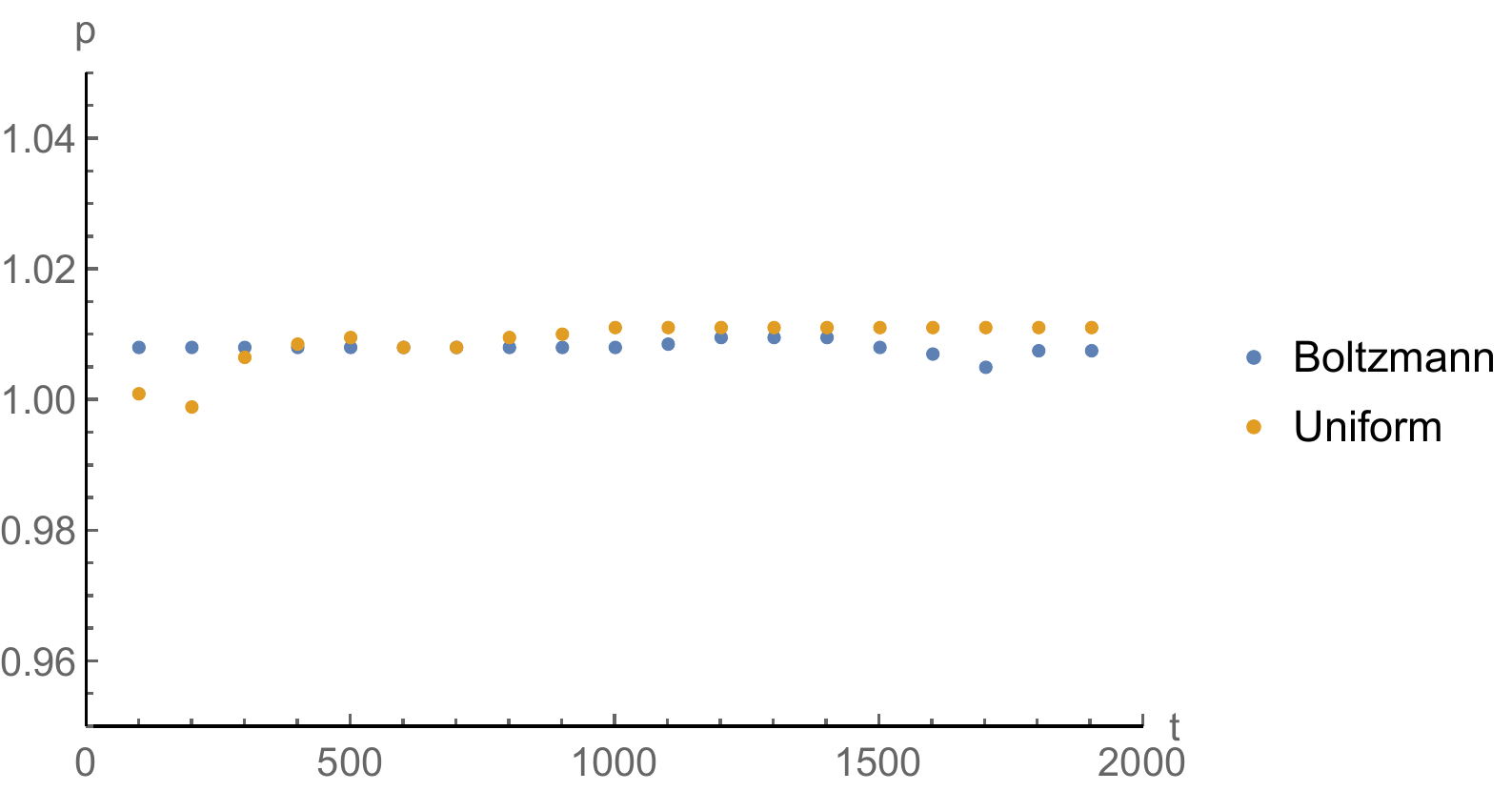}
    \caption{Evolution of the equilibrium price for two differently initialized systems. The parameters are $\bar{a} = \bar{b} = 10$, $\alpha = \beta = 1$, $r = 0.5$, and $\Delta a = 1$. The number of agents $N=1000$.}
    \label{fig:PriceBoltzUnif}
\end{figure}

We consider the distributions found from the simulations after $t=2000$ time steps, at which time we find that the system has reached a steady state. From Figs.~\ref{fig:EvoBoltz} and \ref{fig:EvoUnif} we see that the distribution of goods A and B are strongly and qualitatively dependent on the initial conditions. The distribution of the ratio $b/a$, however, is rather similar for the two initial conditions, which explains the similar supply and demand curves and equilibrium prices (see also Fig.~\ref{fig:PriceBoltzUnif}). It is also found that the agents tend to the ratio $b/a = 1$ in both cases. This implies that the trades an agent makes are strongly dependent on the goods that the agent started with. In order to understand this quantitatively, we calculate below the final distributions within some assumptions that turn out to be reasonable. 

The first assumption is based on the result in Fig.~\ref{fig:PriceBoltzUnif}, which shows the equilibrium price as a function of time steps of both simulations. We find that for both initial conditions, the equilibrium prices remain close to the mean-field price, given by Eq.~(\ref{eq:pMF}). Secondly, we assume the the final distribution function $P^{\rm f} (a,b)$ factorizes, i.e., $P^{\rm f} (a,b)=P^{\rm f}_A (a) P^{\rm f}_B (b)$. Within these assumptions, the final distribution functions are computed as follows. Agents tend to the desired ratio $\gamma(p)$ and to do so, they do $n$ trades involving $\Delta a$ goods of type A and  $p_{\rm MF}$ goods of type B. We allow $n$ to be negative to be able to simultaneously account for buying and selling. In mathematical terms,
\begin{equation}
\begin{aligned}
    a^\textrm{f} &= a^\textrm{i} + n \Delta a, \\
    b^\textrm{f} &= b^\textrm{i} - n p_{\rm MF},
\end{aligned}
\label{eq:ntrades}
\end{equation}

where $a^\textrm{i}$ and $b^\textrm{i}$ are the number of goods of, respectively, type A and B that the agent started out with initially.
We assume that after the trades the ratio of the number of goods is equal to $ (b/a)_{\rm crit} =\gamma(p_{\rm MF})$, and therefore
\begin{equation}
\begin{aligned}
    \frac{b^\textrm{f}}{a^\textrm{f}} &= \frac{b^\textrm{i} - n p_{\rm MF}}{a^\textrm{i} + n \Delta a} = \gamma(p_{\rm MF}), 
\end{aligned}
\label{eq:ratiobfaf}
\end{equation}
from which we find that 
\begin{equation}
\begin{aligned}
    n(a^\textrm{i}, b^\textrm{i}) &= \frac{b^\textrm{i} - \gamma(p_{\rm MF}) a^\textrm{i}}{p_{\rm MF} + \gamma(p_{\rm MF}) \Delta a}.
\end{aligned}
\label{eq:naibi}
\end{equation}
If an agent starts with a number $a\ti$ of goods of the type A, how many goods $b\ti$ of type B would the agent need to end up both in a ratio $b\tf/a\tf = \gamma(p_{\rm MF})$, and at some given final number of good A, $a\tf$? We compute this by inserting Eqs.~(\ref{eq:ratiobfaf}) and (\ref{eq:naibi}) into Eq.~(\ref{eq:ntrades}), so that
\begin{equation}
\begin{aligned}
    a\tf &= a\ti + \frac{b\ti - \gamma(p_{\rm MF}) a\ti}{p_{\rm MF} + \gamma(p_{\rm MF}) \Delta a} \Delta a, \\
    b\ti &= \frac{p_{\rm MF}}{\Delta a} (a\tf - a\ti) + a\tf \gamma(p_{\rm MF}).
\end{aligned}
\end{equation}

We calculate the final probability distribution of good A by adding up the probabilities of the different possibilities of ending up with a certain number $a\tf$:
\begin{equation}
\label{eq:PfAa}
    P\tf_A(a\tf) \propto \int_0^\infty da\ti P\ti\left(a\ti, \frac{p}{\Delta a}(a\tf - a\ti) + a\tf \gamma(p)\right),
\end{equation}
where $P\ti (a,b)$ is the initial distribution of goods. In a very similar way we  find the distribution of good B:
\begin{equation}
    P\tf_B(b\tf) \propto \int_0^\infty db\ti P\ti\left(\frac{\Delta a}{p}(b\tf - b\ti) + \frac{b\tf}{\gamma(p)}, b\ti\right).
\end{equation}
For our initial conditions $P\ti (a,b) = P\ti_A (a) P\ti_B (b)$ and for the Boltzmann-Gibbs distribution $P\ti_A (a) \propto \theta (a) e^{-a/\bar{a}}$, and $P\ti_B (b) \propto \theta (b) e^{-b/\bar{b}}$.
Inserting this into Eq.~(\ref{eq:PfAa}), yields
\begin{equation}
\begin{aligned}
    P\tf_A(a\tf) &\propto \frac{1}{\bar{a}\bar{b}} \int_0^\infty da\ti \exp\left[\left(\frac{p}{\bar{b} \Delta a} - \frac{1}{\bar{a}}\right) a\ti\right] \exp\left\{-\frac{\left[\frac{p}{\Delta a} + \gamma(p)\right]a\tf }{ \bar{b}}\right\} \theta\left[\frac{p}{\Delta a}(a\tf - a\ti) + a\tf \gamma(p)\right].
\end{aligned}
\end{equation}
After integrating and normalizing, we find for the final distribution of good A that 
\begin{equation}
\label{eq:PfAaBoltz}
    P\tf_A(a) = \frac{p + \gamma(p) \Delta a}{p \bar{a} - \Delta a \bar{b}} \left(\exp\left\{-\frac{\left[1 + \frac{\Delta a}{p} \gamma(p)\right] a}{ \bar{a}}\right\} - \exp\left\{-\frac{\left[\frac{p}{\Delta a} + \gamma(p)\right] a}{\bar{b}}\right\} \right).
\end{equation}

\begin{figure}[t]
    \centering
    \includegraphics[width=0.6\textwidth]{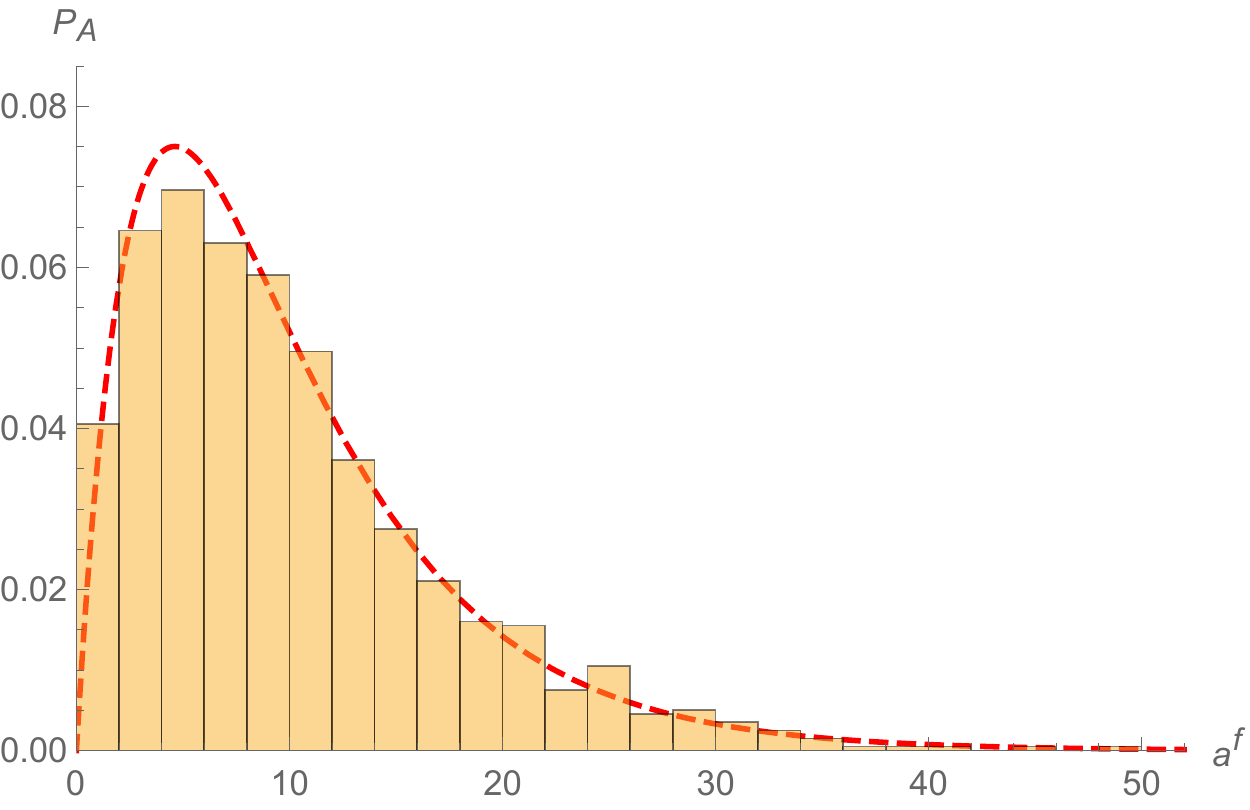}
    \caption{Steady-state distribution as obtained from simulations (bars) of the initially Boltzmann-distributed case with parameters $\bar{a} = 10$, $\bar{b} = 20$, $\alpha = 2$, $\beta = 1$, and $r = 0.5$. The red curve denotes the prediction in Eq.~(\ref{eq:PfAaBoltz}) }
    \label{fig:BoltzCompGen}
\end{figure}

To check this analytical result, we perform a simulation of a system with $N=1000$ agents that are initialized with a Boltzmann-Gibbs distribution of goods. We take the parameters $\bar{a} = 10$, $\bar{b} = 20$, $\alpha = 2$, $\beta = 1$, and $r = 0.5$. Therefore, $p_{\textrm{MF}} = 1$, and $\gamma(p_{\textrm{MF}}) = 2$. Furthermore, $\bar{a}/\alpha = 5 \neq \bar{b}/\beta = 20$, and therefore the mean-field price is not the equilibrium price cf. our earlier discussion. However, according to Fig.~\ref{fig:dAsAlog} it is still a good approximation. Fig.~\ref{fig:BoltzCompGen} shows both the actual final distribution as determined from the simulation, and the predicted distribution from Eq.~(\ref{eq:PfAaBoltz}). It is clear from this figure that the analytical result compares favourably to the results found from the simulations.

\begin{figure}[H]
    \centering
    \includegraphics[width=0.6\textwidth]{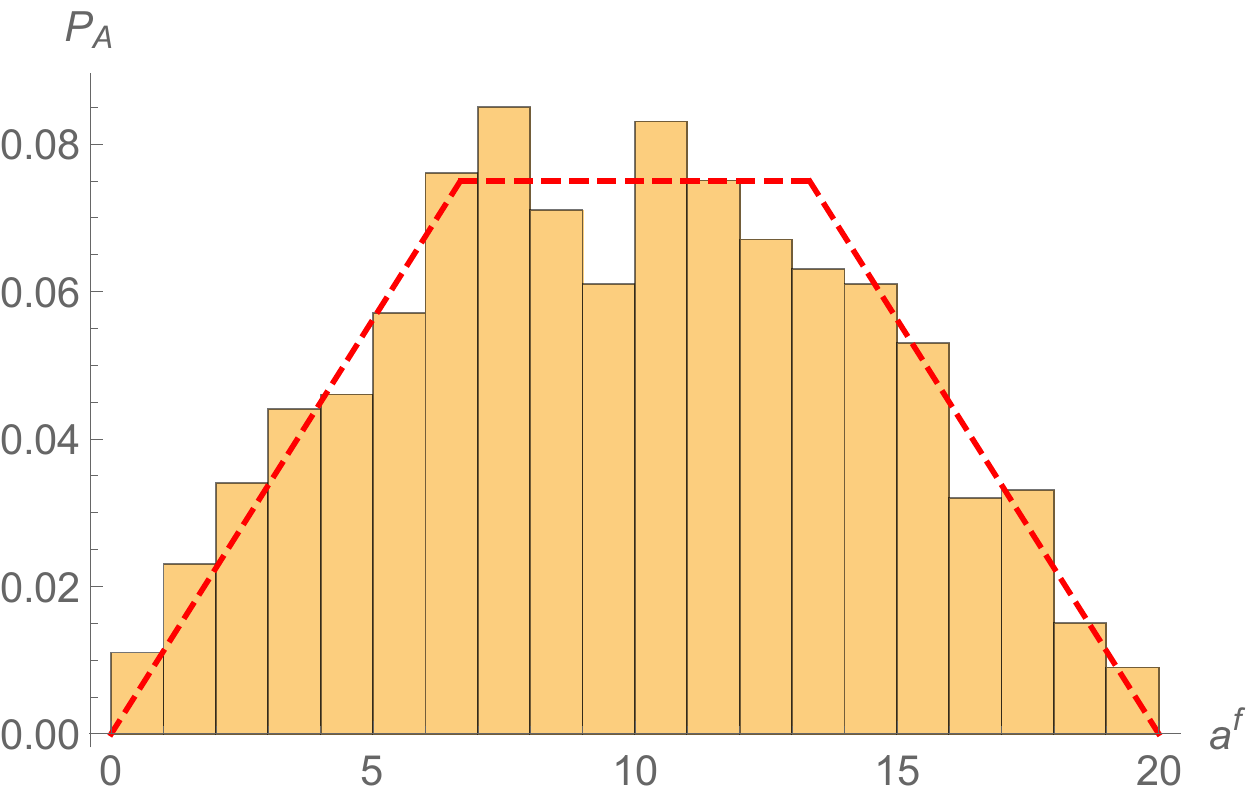}
    \caption{Steady-state distribution as obtained from simulations (bars) of the initially homogeneous distribution with parameters $\bar{a} = 10$, $\bar{b} = 20$, $\alpha = 2$, $\beta = 1$, and $r = 0.5$. The red curve denotes the prediction in Eq.~(\ref{eq:Pequilhomofirst}) }
    \label{fig:UnifCompGen}
\end{figure}

In a similar way we compute the final distribution of goods for a system that is initialized with a uniform distribution of goods, the final result of which is represented as a $2\times2$ matrix contracted with a column and row vector as
\begin{equation}
\label{eq:Pequilhomofirst}
\begin{aligned}
    P\tf_A(a) = \frac{\frac{p}{\Delta a} + \gamma(p)}{4 \bar{a} \bar{b}} &\begin{bmatrix} \theta\Big(\frac{2\bar{a}}{1+\frac{\Delta a}{p}\gamma(p)} - a\Big), & \theta\Big(a - \frac{2\bar{a}}{1+\frac{\Delta a}{p}\gamma(p)}\Big) \end{bmatrix} \cdot  \\
    &\begin{bmatrix} \left[1 + \frac{\Delta a}{p} \gamma(p)\right] a & 2 \frac{\Delta a}{p} \bar{b} \\ 2 \bar{a} & \theta\Big(\frac{2 [p \bar{a} + \Delta a \gamma(p)]}{p + \Delta a \gamma(p)} - a\Big) \Big\{2\bar{a} + 2 \frac{\Delta a}{p} \bar{b} - \left[1 + \frac{\Delta a}{p} \gamma(p)\right]a\Big\} \end{bmatrix} \cdot
     \begin{bmatrix} \theta\Big(\frac{2 \Delta a \bar{b}}{p + \Delta a \gamma(p)} - a\Big) \\ \theta\Big(a - \frac{2 \Delta a \bar{b}}{p + \Delta a \gamma(p)}\Big) \end{bmatrix}.
\end{aligned}
\end{equation}
Using the same parameters as before, we compare this analytical result to the numerical result for the distribution function. This comparison is shown in Fig.~\ref{fig:UnifCompGen}, revealing excellent agreement between the numerical and analytical results.

We conclude that, for the type of dynamics we considered, the final steady-state distributions depend strongly on the initial conditions. This is understood as follows: Because of the approximately constant price, the agents are not able to buy at a low price and sell at a high price, or vice versa. Therefore they cannot move up or down in 'wealth', i.e., in number of goods, and they are stuck with the wealth they were given initially. The dynamics then result in the individual agents trading such that their ratio of numbers of goods becomes close to the ideal ratio as determined by the utility.

\section*{Conclusion, Discussion and Outlook}
In this article, we have developed a simple statistical-physics-inspired model for a market of goods in which intrinsic statistical fluctuations, engendered by the distribution of goods over the agents, drive the demand and supply. We found that the equilibrium price depends on the distribution of goods over the agents and is generically different from the equilibrium price that results when all agents have an equal number of goods. This difference is largest when the price and number of goods that are traded and the distribution of goods over the agents are such there is a large fraction of agents can neither buy nor sell. In addition, our model provides a straightforward way to compute intrinsic fluctuations in the equilibrium price. 

We considered the total number of goods per type to be conserved quantities, in analogy to conserved quantities in physics and motivated by the success of the model by Yakovenko and co-workers \cite{yakovenko}. The assumption of conserved goods leads to the so-called Boltzmann-Gibbs distribution for the goods over the agents. In the case that such a conservation law is lacking, or when different constraints are present, one might consider to construct the probability distribution for the goods over the agents from the maximum entropy principle \cite{PhysRev.106.620}, \cite{Scharfenaker2020}. In fact, the Boltzmann-Gibbs distribution is a specific case of a probability distribution that obeys this principle. A crucial ingredient of our model is the utility for the goods that the agents have, which is identical for all agents. From a statistical-physics perspective, the utility, however, has, to the best of our knowledge, no analogue, despite the thermodynamic analogues that were put forward by Saslow \cite{doi:10.1119/1.19110}.

The difference with physical systems is also apparent in the simple dynamics that we considered for our model. The dynamics is inspired by economic trade at a market after setting an equilibrium price. While perhaps reasonable from an economic point-of-view, this type of dynamics does not obey detailed balance and, as a result, the steady-state distribution functions strongly depend on the initial conditions. 

Possible generalizations of our model could consider a larger number of different types of commodities and/or differences in utilities between the agents. This would probably make the dynamics less restrictive and more amendable to analogies with dynamics of physical systems. Our long-term goal of constructing the model presented in this article is to use it to build towards a macroscopic description, viz. the transition from statistical physics to thermodynamics. Notwithstanding the above-mentioned incongruities, there are some possible inroads towards this goal. A first step, for example, is to connect two systems described by the model that we proposed here, and let them interact and equilibrate by exchange of goods.  In future work we intend to explore such generalizations. 

RvR and RD are members of the D-ITP consortium, a program of the Netherlands Organisation for
Scientific Research (NWO) that is funded by the Dutch
Ministry of Education, Culture and Science (OCW).

\bibliography{main}

\section*{Author contributions statement}

RD and RvR conceived the project, and JM carried out all calculations and wrote the first version of the manuscript. All authors contributed to the manuscript. 

\section*{Additional information}

The authors have no competing interests. 

\end{document}